\documentclass[11pt]{article}

\usepackage[margin=1in]{geometry}
\usepackage{graphicx}
\usepackage{amsmath,amssymb}
\usepackage{booktabs}
\usepackage{tabularx}
\usepackage{threeparttable}
\usepackage{multirow}
\usepackage{caption}
\usepackage{subcaption}
\usepackage{placeins}
\usepackage{xcolor}
\usepackage{enumitem}
\usepackage[hidelinks]{hyperref}
\usepackage[numbers,super,sort&compress]{natbib}
\usepackage{bm}
\newcommand{\vcond}{v_{\mathrm{cond}}}

\graphicspath{{./Final_Graphs/}}

\title{Degrees-of-Freedom Approximations for Conditional-Mean Inference in Random-Lot Stability Analysis}

\author{
Andrew T. Karl$^{1}$ \quad Heath Rushing$^{2}$ \quad Richard K. Burdick$^{3}$ \quad Jeff Hofer$^{4}$\\[0.6ex]
\small $^{1}$Karl Statistical Services LLC, Aurora, CO, 80016, USA. \texttt{akarl@asu.edu}. ORCID: \href{https://orcid.org/0000-0002-5933-8706}{0000-0002-5933-8706}.\\
\small $^{2}$Adsurgo LLC, San Antonio, TX, 78232, USA. \texttt{heath.rushing@adsurgo.com}.\\
\small $^{3}$Burdick Statistical Consulting, LLC, Colorado Springs, CO, 80923, USA. \texttt{rickbasu@aol.com}.\\
\small $^{4}$JDH CMC Stats, LLC, Avon, IN, 46123, USA. \texttt{jeffreyhofer@jdhcmcstats.com}.\\
}

\date{}

\begin{document}
\maketitle

\begin{abstract}
Linear mixed models are widely used for pharmaceutical stability trending when sufficient lots are available. Expiry support is often based on whether lot-specific conditional-mean confidence limits remain within specification through a proposed expiry, and these limits depend on the denominator degrees-of-freedom (DDF) method used for $t$-based inference. We document an operationally important boundary-proximal phenomenon: when a fitted random-effect variance component is close to zero, Satterthwaite DDF---and, in sensitivity analyses, Kenward--Roger DDF---for conditional-mean predictions can collapse, inflating $t$ critical values and producing very wide and sometimes nonmonotone pointwise confidence limits on scheduled time grids. In the settings studied here, containment DDF yields stable degrees of freedom and avoids sharp discontinuities as variance components approach the boundary. In a balanced random-intercept design, a closed-form calculation gives the conditional prediction variance and delta-method Satterthwaite DDF, showing why DDF can be extremely small near the boundary when the prediction point is near the design centroid. A worked example and simulation studies show that DDF choice can materially change pass/fail conclusions even when the observed measurements themselves are not close to specification limits. Containment-based inference with the full random-effects model provides a single framework without model switching. When containment is unavailable, a 10\% variance-contribution reduction procedure mitigates extreme Satterthwaite behavior by simplifying only when fitted contributions at the proposed expiry are negligible. AICc step-down is best treated as a sensitivity analysis because it can be too aggressive when the proposed-expiry margin is small.
\end{abstract}

\section{Introduction}

Regression-based analysis underpins pharmaceutical stability evaluation, including shelf-life (or retest period) estimation, investigation of stability-indicating trends, and ongoing product lifecycle trending.
ICH Q1E describes a fixed-lot ANCOVA framework for analyzing stability data and pooling across lots when appropriate, and it provides guidance for handling one-sided and two-sided acceptance criteria.\citep{ICHQ1E2003} We present the lower-sided case; upper-sided and two-sided criteria use the corresponding upper or paired bounds, with the same DDF mechanism.
Recent Step~2a/b draft ICH Q1 guidance continues to emphasize model-based evaluation of stability profiles and places additional emphasis on mixed models as an analysis option when sufficient lots are available.\citep{ICHQ1Draft2025}
However, the draft guidance is silent on the choice of denominator degrees of freedom for mixed-model inference, leaving analysts to rely on software defaults that differ across platforms.\citep{ICHQ1Draft2025}

A common stability decision rule is based on confidence limits for a mean regression line: expiry is the time at which an appropriate confidence limit intersects an acceptance criterion.\citep{ICHQ1E2003,ICHQ1Draft2025}
In fixed-lot analyses, this is often implemented by fitting a lot-specific ANCOVA model and selecting the worst-case lot (earliest intersection), with pooling guided by Q1E-recommended $p=0.25$ cutoffs for slope and intercept pooling.\citep{ICHQ1E2003}
Because this fixed-lot procedure remains standard in stability analysis, we include a Q1E-style fixed-lot comparator as a baseline alongside pooled OLS and random-lot mixed-model approaches.

In random-lot analyses, an analogous approach evaluates lot-specific conditional-mean lower confidence limits derived from empirical best linear unbiased predictors (EBLUPs).
This conditional approach targets a practical question: given the observed lots, do the lot-specific conditional-mean lower confidence limits remain above specification through a proposed expiry on a scheduled grid?

Inference in linear mixed models requires approximations for the reference distribution used with estimated standard errors of contrasts.
Standard software offers multiple denominator degrees-of-freedom (DDF) methods for $t$- and $F$-based inference; in the SAS/JMP workflows motivating this paper, the routine options are containment (CONTAIN) and Satterthwaite (SAT).\citep{SASMixedDDFM152}
In stability applications the DDF method directly affects conditional-mean confidence limits and can change expiry/extension support conclusions.
In particular, we document a boundary-proximal instability for conditional-mean predictions: very small positive fitted random-effect variance components can cause variance-component-dependent DDF methods to collapse, inflating the $t$ multiplier and widening pointwise confidence limits; Section~\ref{sec:vcond_mechanism} gives the mechanism.
Kenward--Roger (KR) is also variance-component- and contrast-dependent; in the worked example and appendix mechanism plots it follows the same near-boundary pattern as SAT rather than resolving the collapse.

At the worked-example profiler point shown in Figures~\ref{fig:jmp-default}--\ref{fig:sas-overlay} (Lot~G at Month~24), \texttt{PROC MIXED} reports a conditional-mean Satterthwaite DDF of 1 (Appendix~\ref{app:satt_rebuild}). The corresponding one-sided 95\% $t$ critical value is therefore $t_{0.95,1}=6.31$. For the same random-intercept fit, containment reports DDF$=111$, giving $t_{0.95,111}\approx1.66$. KR also reports conditional DDF$=1$ at this point and therefore uses the same $t$ multiplier as SAT; because the KR-adjusted prediction standard error is larger, the KR interval is wider than the SAT interval. This example illustrates that, near the boundary, variance-component-dependent methods can materially inflate pointwise confidence-limit widths because of the DDF approximation (and, for KR, the associated prediction-variance adjustment), not because the observed data are inadequate.

We contrast this behavior with containment, which yields stable DDF and avoids sharp discontinuities as variance components approach the boundary at zero.
Containment is available in SAS \texttt{PROC MIXED} and in JMP Pro (version 19 and later), but it is not available in base JMP where Satterthwaite is the default for conditional-mean inference.\citep{SASMixedDDFM152,JMPMixedModelOptions19,JMPKH}
Containment-based inference with the full random-effects model provides a single modeling framework, avoiding the discontinuities that arise from data-dependent model reduction or selection rules---whether Q1E $p=0.25$ pooling tests, variance-contribution thresholds, or information-criterion step-down rules.
Because some stability analysts work in SAT-only environments, we also evaluate two pragmatic SAT-only approaches: (i) a simple 10\% variance-contribution reduction procedure that drops random-effect terms only when their fitted contribution at the proposed expiry is negligible, and (ii) an AICc step-down among candidate random-lot and pooled models.
These procedures can mitigate extreme near-boundary SAT behavior, but they introduce their own discontinuities through data-dependent model simplification; AICc step-down can also be too aggressive in expiry settings close to the specification limit.

Section~2 defines the mixed-model setup, conditional-mean confidence limits, DDF methods, the analytical near-boundary mechanism, and the reduction and fixed-lot comparators.
Section~3 presents the worked example and simulation results, including decision operating characteristics at the proposed expiry.
Section~4 discusses implications for routine stability analytics and provides practical recommendations.
Section~5 concludes. The appendices give the known-variance decision reference, the worked-example SAT/KR diagnostics, the closed-form balanced-design calculation, and supplementary sensitivity analyses.

\section{Methods}

\subsection{Model and conditional-mean confidence bounds}

Let $Y_{ij}$ denote the measured attribute for lot $i$ at time $t_{ij}$ (months). We consider the random-lot stability linear mixed model with a random intercept and random slope (Model~(1))
\begin{equation}
Y_{ij} = \beta_0 + \beta_1 t_{ij} + b_{0i} + b_{1i} t_{ij} + \varepsilon_{ij},
\label{eq:lmm}
\end{equation}
where $(\beta_0,\beta_1)$ are fixed effects, $(b_{0i},b_{1i})$ are lot-level random effects, and $\varepsilon_{ij}\sim N(0,\sigma_e^2)$ are independent residuals.
We assume mean-zero Normal random effects with $\mathrm{Var}(b_{0i})=\sigma_{b0}^2$, $\mathrm{Var}(b_{1i})=\sigma_{b1}^2$, and $\mathrm{Cov}(b_{0i},b_{1i})=0$. This diagonal random-effects structure is a pragmatic stability-modeling simplification used to keep the fitted covariance structure identifiable with modest lot counts; allowing an intercept--slope covariance is possible with richer data but does not remove the near-boundary variance-component issue studied here.

The random-intercept model (Model~(2)) is
\begin{equation}
Y_{ij} = \beta_0 + \beta_1 t_{ij} + b_{0i} + \varepsilon_{ij},
\label{eq:ri}
\end{equation}
obtained from Model~(1) by setting $b_{1i}\equiv 0$.
The pooled OLS model (Model~(3)) is
\begin{equation}
Y_{ij} = \beta_0 + \beta_1 t_{ij} + \varepsilon_{ij},
\label{eq:ols}
\end{equation}
obtained by setting $b_{0i}\equiv b_{1i}\equiv 0$.

For lot $i$ at time $t$, the conditional mean is
\[
\mu_i(t)=\beta_0+\beta_1 t + b_{0i} + b_{1i}t,
\]
estimated using EBLUPs as
\[
\widehat{\mu}_i(t)=\widehat{\beta}_0+\widehat{\beta}_1 t + \widehat{b}_{0i} + \widehat{b}_{1i}t.
\]

For lower-sided acceptance criteria (as in Q1E\citep{ICHQ1E2003}), we form one-sided $100(1-\alpha)\%$ lower confidence limits for $\mu_i(t)$ as
\begin{equation}
\mathrm{LCL}_i(t)=\widehat{\mu}_i(t) - t_{1-\alpha,\ \nu(t,i)} \cdot \widehat{\mathrm{se}}\{\widehat{\mu}_i(t)\},
\label{eq:lcl}
\end{equation}
where $\nu(t,i)$ is an approximate denominator degrees of freedom that may depend on the contrast defining $\widehat{\mu}_i(t)$.
We refer to the collection of these limits over a time grid as pointwise lower confidence limits (informally, a pointwise band); these are pointwise, not simultaneous, confidence limits.
Throughout we use $\alpha=0.05$ (one-sided 95\%).
In \texttt{PROC MIXED}, Eq.~\eqref{eq:lcl} is implemented by requesting two-sided 90\% confidence limits for predictions (\texttt{ALPHAP=0.10}) and retaining the lower limit.

For an upper-sided acceptance criterion, the analogous pointwise upper confidence limit is obtained by replacing the minus sign in Eq.~\eqref{eq:lcl} with a plus sign and comparing to the upper specification limit.
For a two-sided criterion, both the lower and upper limits are evaluated.
The boundary-proximal DDF instability discussed in this paper is symmetric with respect to the side of the specification limit: the side changes the direction of the confidence bound, but the standard error and DDF calculation for the conditional-mean contrast are unchanged.
Thus upper-sided and two-sided criteria inherit the same potential SAT DDF collapse when the relevant conditional-mean prediction has a near-boundary fitted variance component.

For Model~(2) we summarize the lot variance contribution by
\begin{equation}
\mathrm{vcfrac} = \frac{\sigma_{b0}^2}{\sigma_{b0}^2+\sigma_e^2}.
\label{eq:vcfrac}
\end{equation}
In the balanced random-intercept calculation of Section~\ref{sec:vcond_mechanism}, the analytic variance fraction $p=\tau/(\tau+s)$ is the same quantity as $\mathrm{vcfrac}$, written with $\tau=\sigma_{b0}^2$ and $s=\sigma_e^2$.
For Model~(1), the lot-intercept and lot-slope contributions to the marginal variance at a given time $t_*$ are denoted $p_{b0}(t_*)$ and $p_{b1}(t_*)$, respectively; these time-dependent analogs of $\mathrm{vcfrac}$ are defined in Section~\ref{sec:reduced} and used only as part of the pragmatic reduction rule described there.

\subsection{Specification compliance and first-crossing time}
\label{sec:oc}

At time $t$, let $\mathrm{LCL}_i(t)$ denote the lower confidence limit from Eq.~\eqref{eq:lcl} for lot $i$.
Let $\mathrm{LSL}$ denote the lower specification limit (acceptance criterion).
For lower-sided specifications, compliance is scored by the signed quantity $\mathrm{LCL}_i(t)-\mathrm{LSL}$; specification compliance fails at $t$ if this quantity is negative.

To align with worst-case-lot shelf-life logic, we summarize each lot by its first-crossing time on the evaluation grid, defined as the earliest month with $\mathrm{LCL}_i(t)-\mathrm{LSL}<0$.
If no crossing occurs on the grid, the lot is treated as not crossing through the last evaluated month.
Within a dataset, the worst-case first-crossing month is the minimum first-crossing time across lots among lots that cross on the grid; if no lot crosses, the dataset is treated as supporting the criterion through the last evaluated month.

Nonmonotone pointwise limits could potentially induce nonmonotone compliance on a scheduled grid (failure followed by later satisfaction of the lower-sided criterion). We note this qualitative possibility, but we do not attempt to quantify it in the simulation study.

\subsection{Inference approaches and DDF settings}
\label{sec:methods_ddf}
\label{sec:satt_cond_vs_marg}

The mixed-model conditional-mean lower confidence limits in Eq.~\eqref{eq:lcl} depend on both the estimated standard error and the $t$ critical value, and, through the latter, on the denominator degrees-of-freedom (DDF) approximation.
In the SAS/JMP software motivating this paper, commonly available DDF options for mixed models are containment (CONTAIN) and Satterthwaite (SAT).\citep{SASMixedDDFM152}
The containment DDF is design-based and fixed across contrasts, whereas the Satterthwaite DDF is contrast-specific and depends on fitted variance components.
Consequently, the Satterthwaite DDF can differ across lots and across time points within a single fitted model, so the $t$ critical value used for a given lot at a given month need not equal the value used for another lot or another month.

Satterthwaite DDF values are obtained by matching moments of the estimated prediction variance to a scaled chi-squared reference.
When a variance component is close to zero, this moment-matching calculation can yield very small DDF for some conditional-mean contrasts.
The Kenward--Roger method is also variance-component- and contrast-dependent.\citep{KenwardRoger1997,KenwardRoger2009,SASMixedDDFM152}
We evaluate Kenward--Roger as an additional appendix sensitivity analysis for the worked example and for the margin/DDF mechanism plots (Appendix~\ref{app:additional_sensitivity}); here and in Appendix~\ref{app:kr_sensitivity}, KR refers to \texttt{PROC MIXED} \texttt{DDFM=KR2} output.

We compare the following analysis approaches (Table~\ref{tab:approaches}): pooled OLS (Model~(3)); a fixed-lot comparator (FIXED); and mixed-model inference using CONTAIN and SAT.
For SAT-only environments, we also evaluate two procedures intended to mitigate near-boundary SAT behavior: a 10\% variance-contribution reduction procedure (SAT\_reduced) and an AICc step-down among candidate random-lot and pooled models (SAT\_AICc).
The appendix also includes a p-value reduction sensitivity comparator (SAT\_covtest25).

\begin{table}[htbp]
\centering
\caption{Analysis approaches compared in this study.}
\label{tab:approaches}
\small
\begin{tabular}{@{}p{0.15\textwidth} p{0.49\textwidth} p{0.27\textwidth}@{}}
\toprule
Method & Inference & Notes \\
\midrule
OLS & Pooled regression (Model~(3)) & Matches the data-generating model when $\sigma_{b0}^2=\sigma_{b1}^2=0$ \\
FIXED & Lot and lot-by-time fixed effects & Q1E-style comparator \citep{ICHQ1E2003} \\
CONTAIN & Mixed model with \texttt{DDFM=CONTAIN} & Fixed DDF; stable at the boundary \\
SAT & Mixed model with \texttt{DDFM=SAT} & Contrast-specific DDF; can collapse near the boundary \\
SAT\_reduced & 10\% variance-contribution reduction procedure, then compute limits from the resulting model using SAT or OLS as applicable & Mitigates extreme SAT behavior in SAT-only environments \\
SAT\_AICc &
AICc step-down among Models~(1)--(3), then compute limits using SAT or OLS as applicable &
Automated step-down; can be too aggressive in some near-boundary expiry settings \\
SAT\_covtest25 &
\texttt{COVTEST} $p>0.25$ step-down, then compute limits using SAT or OLS as applicable &
Appendix sensitivity comparator, not shown in the main mechanism figures; natural Q1E-style p-value rule, but non-regular at the variance-component boundary \\
\bottomrule
\end{tabular}
\end{table}

For a predicted mean, let $\widehat v$ denote its estimated prediction variance,
$\widehat v=\widehat{\mathrm{se}}\{\widehat\mu_i(t)\}^2$, the value reported by
\texttt{PROC MIXED} as \texttt{StdErrPred}$^2$.
We distinguish the conditional prediction variance $\widehat v_{\mathrm{cond},i}(t)$, used for the lot-specific EBLUP-based prediction $\widehat\mu_i(t)$, from the marginal prediction variance $\widehat v_{\mathrm{marg}}(t)$, used for the fixed-effect-only mean $x(t)^\top\widehat\beta$.
The conditional variance is the object of interest for the worst-case-lot expiry question because the limits are formed for observed lots and the prediction includes the fitted lot effect.
The approximate Satterthwaite DDF uses the moment-matching form\citep{Satterthwaite1946,GiesbrechtBurns1985,SASMixedDDFM152}
\begin{equation}
\nu \;=\; \frac{2\,\widehat v^{\,2}}{\widehat{\operatorname{Var}}(\widehat v)},
\qquad
\widehat{\operatorname{Var}}(\widehat v)\;\approx\;
g(\widehat\theta)^\top \,\widehat\Omega\, g(\widehat\theta),
\label{eq:satt_df_delta}
\end{equation}
where $\theta$ collects the covariance parameters, $\widehat\Omega$ is their
estimated asymptotic covariance (the \texttt{AsyCov} table in \texttt{PROC MIXED}),
$v(\theta)$ is the prediction variance for the given lot--time contrast regarded as a function of $\theta$ (so that
$\widehat v=v(\widehat\theta)$), and $g(\theta)=\partial v(\theta)/\partial\theta$ is its gradient.
The conditional quantity $\widehat v_{\mathrm{cond},i}(t)$ depends on the joint mixed-model solution for the fixed and random effects (blocks of the inverse coefficient matrix in the mixed-model equations), whereas $\widehat v_{\mathrm{marg}}(t)$ depends only on $\widehat{\operatorname{Cov}}(\widehat\beta)$.
When one or more fitted variance components are near the boundary at zero, the resulting gradient $g(\widehat\theta)$ can have much larger magnitude for conditional predictions than for marginal ones, as in the worked example below.

Appendix~\ref{app:satt_rebuild} reconstructs these quantities at the worked-example
profiler point used in Figures~\ref{fig:jmp-default}--\ref{fig:sas-overlay} and shows that,
within the delta-method formula, the amplified conditional gradient is the driver of the DDF collapse.

\subsection{Mechanism: concavity and saturation of the conditional prediction variance}
\label{sec:vcond_mechanism}

The boundary-proximal DDF collapse can be understood by examining the prediction-variance function that enters Eq.~\eqref{eq:satt_df_delta}.
For a conditional mean of an observed lot, this function is nonlinear in the random-effect variance component: it has an OLS floor at the boundary and a finite large-variance limit.
The Satterthwaite calculation then maps uncertainty in the plug-in prediction variance into a denominator degrees-of-freedom value through the moment-matching ratio \(2\widehat v^2/\widehat{\operatorname{Var}}(\widehat v)\). Near the $\tau=0$ boundary, covariance-parameter uncertainty is propagated through a steep part of this bounded, concave function, making \(\widehat{\operatorname{Var}}(\widehat v)\) large relative to \(\widehat v^2\).

\paragraph{Balanced random-intercept calculation.}
Consider the balanced random-intercept design in which Model~(2) is fit to $L$ lots, each observed at the same $m$ time points, and the target is the lot-specific conditional mean for one observed lot at time $t_*$.
Write $\tau=\sigma_{b0}^2$, $s=\sigma_e^2$, $x_*=(1,t_*)^\top$, and let $X$ denote the full fixed-effect design matrix for all $Lm$ observations.
Appendix~\ref{app:vcond_derivation} shows from the mixed-model equations that the conditional prediction-error variance for $\widehat\mu_i(t_*)-\mu_i(t_*)$ is
\begin{equation}
\vcond(t_*;\tau,s)
=
s\,x_*^\top(X^\top X)^{-1}x_*
+
\frac{s\tau(L-1)}{L(s+m\tau)}.
\label{eq:vcond_closed}
\end{equation}
The first term is the fixed-effect leverage component of the prediction-error variance. The second term accounts for uncertainty in predicting the observed lot's random intercept after shrinkage. Appendix~\ref{app:vcond_derivation} derives this decomposition by rotating the lot effects into their average direction and the remaining $L-1$ orthogonal contrast directions.

For compactness, write
\[
A_* = x_*^\top(X^\top X)^{-1}x_*.
\]

Equation~\eqref{eq:vcond_closed} has an OLS floor at the boundary,
\[
\vcond(t_*;0,s)=sA_*,
\]
and a finite large-variance limit,
\[
\lim_{\tau\to\infty}\vcond(t_*;\tau,s)=sA_*+\frac{s(L-1)}{Lm}.
\]
Its derivatives with respect to the lot variance component are
\begin{equation}
\frac{\partial \vcond}{\partial \tau}
=
\frac{s^2(L-1)}{L(s+m\tau)^2},
\qquad
\frac{\partial^2 \vcond}{\partial \tau^2}
=
-\frac{2ms^2(L-1)}{L(s+m\tau)^3}<0.
\label{eq:vcond_derivatives}
\end{equation}
Thus $\vcond$ is increasing but concave in $\tau$, with the steepest finite slope at the boundary and decreasing sensitivity as $\tau$ grows.
The companion derivative with respect to the residual variance, holding $\tau$ fixed, is
\begin{equation}
\frac{\partial \vcond}{\partial s}
=
A_*+
\frac{m(L-1)\tau^2}{L(s+m\tau)^2},
\label{eq:vcond_derivative_s}
\end{equation}
which reduces to $A_*$ at the boundary.

The lot-contrast term has the standard balanced-BLUP shrinkage form,
\[
\frac{s\tau}{s+m\tau}
=
\frac{s}{m}\frac{m\tau}{s+m\tau},
\]
where $m\tau/(s+m\tau)$ is the shrinkage weight.\citep{robinson1991blup,SearleCasellaMcCulloch1992}
Because the reduction rule and simulations are expressed in terms of the variance fraction, we quantify the midpoint of the rise from the OLS floor to the large-variance limit on that scale.
Let
\[
p=\frac{\tau}{\tau+s},
\qquad
c=\frac{L-1}{L},
\qquad
D(p)=1+(m-1)p.
\]
Because $\tau=sp/(1-p)$,
\begin{equation}
\vcond(t_*;s,p)
=
s\left\{A_*+c\frac{p}{D(p)}\right\}.
\label{eq:vcond_p}
\end{equation}
Holding $p$ fixed for the derivative with respect to $s$,
\begin{equation}
\frac{\partial \vcond}{\partial s}
=
A_*+c\frac{p}{D(p)},
\qquad
\frac{\partial \vcond}{\partial p}
=
\frac{sc}{D(p)^2},
\qquad
\frac{\partial^2 \vcond}{\partial p^2}
=
-\frac{2sc(m-1)}{D(p)^3}<0.
\label{eq:vcond_p_derivatives}
\end{equation}
The fraction of the possible rise from the OLS floor to the large-variance limit that has occurred by variance fraction $p$ is
\begin{equation}
\frac{\vcond(t_*;s,p)-\vcond(t_*;s,0)}
{\lim_{q\uparrow1}\vcond(t_*;s,q)-\vcond(t_*;s,0)}
=
\frac{mp}{1+(m-1)p}.
\label{eq:vcond_fraction_rise}
\end{equation}
Setting this fraction equal to one half gives
\[
p^*=\frac{1}{m+1}.
\]
For the worked-example design, $m=9$, so $p^*=0.10$.
At this point the curve has traversed one half of its possible rise, and the local slope with respect to $p$ has fallen to $1/\{1+8(0.10)\}^2=0.31$ of its boundary value.
The fitted worked-example value is $\widehat p=0.004491/(0.004491+0.2360)=0.0187$, placing the scoring point well inside the steep near-boundary part of the curve.

Appendix~\ref{app:vcond_derivation} combines Eqs.~\eqref{eq:vcond_derivatives}--\eqref{eq:vcond_derivative_s} with the closed-form balanced-design covariance matrix of $(\hat\tau,\hat s)$ to give the corresponding analytic delta-method Satterthwaite DDF.
For the worked-example design and the Lot~G / Month~24 scoring point, the right-hand boundary limit is $\nu_\Delta(0^+)\approx0.071$, whereas evaluating the same expression at $\hat\tau=0.004491$ gives $\nu_\Delta\approx0.84$. This is the continuous \texttt{AsyCov}-based reconstruction in Appendix~\ref{app:satt_rebuild}; \texttt{PROC MIXED} reports the operational DDF as 1.00.
The collapse is not only a boundary-limit phenomenon. The delta-method DDF $\nu_\Delta$ rises from about 0.071 at $p=0^+$ to only about 0.84 at $\widehat p=0.0187$, both far below the within-lot residual degrees of freedom $d_W=111$ approached in the large-variance limit. The small DDF reflects first-order propagation of covariance-parameter uncertainty through a bounded, concave prediction-variance function, evaluated where the local slope is still near its boundary maximum while the function value remains far below its large-variance limit.

\begin{figure}[htbp]
\centering
\includegraphics[width=0.96\linewidth]{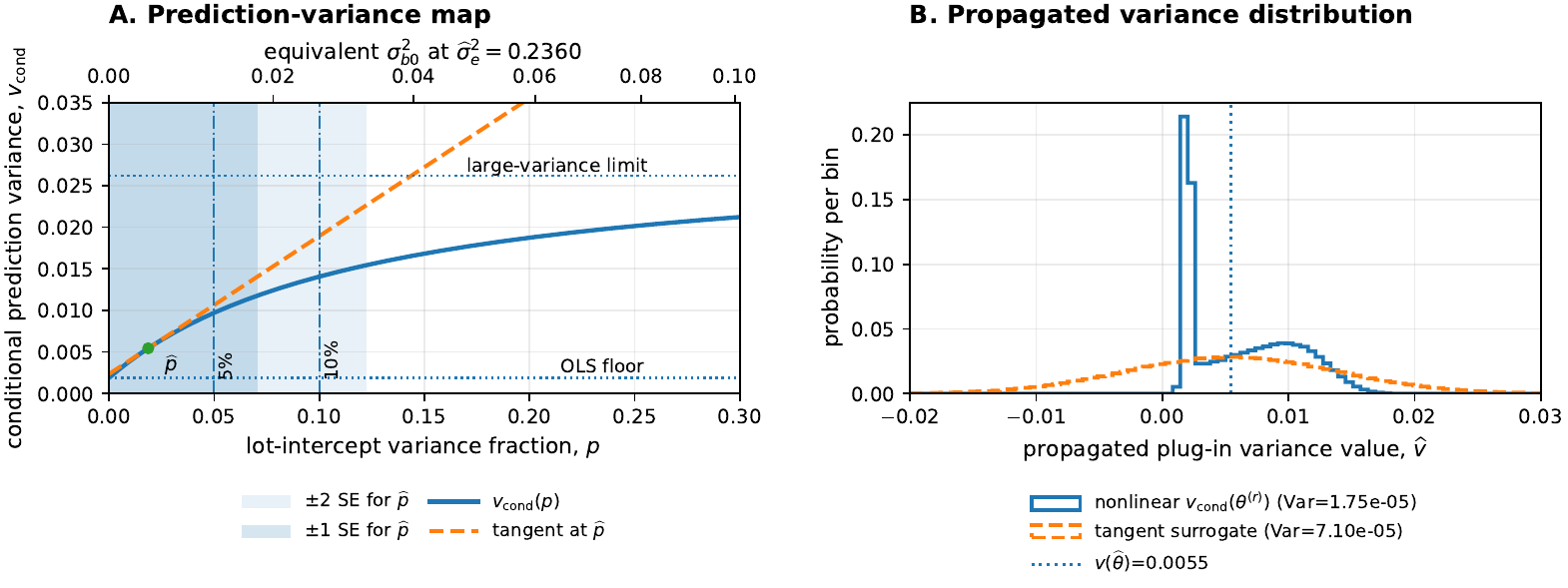}
\caption{Mechanism of the Satterthwaite DDF collapse in the worked example. Panel A shows $v_{\mathrm{cond}}(p)$ with $\sigma_e^2$ fixed at $\widehat\sigma_e^2$. Panel B propagates the covariance-parameter working uncertainty $N(\widehat\theta,\widehat\Omega)$ through the bounded nonlinear map and through the first-order tangent surrogate.}
\label{fig:vcond_concavity}
\end{figure}

Figure~\ref{fig:vcond_concavity} visualizes the mechanism in two steps.
Panel A shows a one-dimensional slice of the prediction-variance map, holding $s=\widehat s$ fixed and varying the variance fraction $p$. The solid curve is the closed-form conditional prediction-error variance, and the dashed line is the local tangent used by the first-order delta method at the fitted value. Panel B uses the corresponding two-parameter map in $\theta=(\tau,s)$ and shows the implication for the Satterthwaite denominator.

In Panel B we draw $M=500{,}000$ values
\[
\theta^{(r)}=(\tau^{(r)},s^{(r)})^\top
\sim
N(\widehat\theta,\widehat\Omega),
\]
using the covariance-parameter \texttt{AsyCov} matrix from Appendix~\ref{app:satt_rebuild}. The solid histogram evaluates the bounded nonlinear map $\vcond(t_*;\tau^{(r)},s^{(r)})$, with nonpositive $\tau^{(r)}$ values mapped to the OLS floor $s^{(r)}A_*$. The dashed histogram evaluates the unrestricted first-order tangent surrogate
\[
L\{\theta^{(r)}\}
=
\vcond(t_*;\widehat\theta)
+
g(\widehat\theta)^\top\{\theta^{(r)}-\widehat\theta\}.
\]
Unlike the bounded nonlinear map, the tangent surrogate is evaluated at each $\theta^{(r)}$ with no positivity constraint imposed on the perturbations or on the resulting $\widehat v$. This is deliberate: the variance of this unrestricted linear approximation is exactly $g(\widehat\theta)^\top\widehat\Omega g(\widehat\theta)$, the denominator component used by the first-order Satterthwaite calculation. Thus, any negative tail values should not be interpreted as feasible prediction variances, but as a feature of the unrestricted linear approximation used to compute the SAT denominator.

Panel B is therefore not a refitting bootstrap and is not proposed as an operational correction. It is a diagnostic, in the spirit of bootstrap and Monte Carlo checks for mixed-model prediction intervals, for visualizing how the unreduced SAT denominator is generated.\citep{ChatterjeeLahiriLi2008}
In the worked example, the tangent surrogate has propagated variance $7.10\times10^{-5}$ (Monte Carlo estimate; the analytic delta-method value $g^\top\widehat\Omega g$ is $7.07\times10^{-5}$, Table~\ref{tab:satt_rebuild} and Appendix~\ref{app:closedform}), whereas direct evaluation of the bounded nonlinear map gives $1.75\times10^{-5}$. With the same plug-in numerator $2\widehat v^2$, these correspond to DDF values of 0.84 and 3.40, respectively; the former is the software-reconstructed SAT mechanism, and the latter is reported only to quantify the linear-versus-nonlinear propagation gap and is not proposed as an alternative DDF.
Thus, for routine conditional-mean expiry analyses, the practical remedies considered here remain containment DDF when available and a pre-specified SAT-reduced model when SAT is the only available DDF option.
KR is evaluated empirically in the worked example and appendix sensitivity analyses because the KR-adjusted prediction-variance and DDF ingredients for conditional EBLUP-based predictions are not exposed in the same reconstruction-ready form as the Satterthwaite \texttt{AsyCov} calculation.

\subsection{Variance-contribution reduction rule (reduced variant)}
\label{sec:reduced}

When estimated variance components are very small, a common practical response is to simplify the random-effects structure and refit a reduced model.
We implement a variance-contribution threshold rule consistent with this practice and with published discussions of practical significance for variance components in pharmaceutical settings (e.g., Burdick and Ermer\citep{BurdickErmer2019}).
Burdick and Ermer discuss a 20\% practical-significance threshold in the context of precision of a reportable value and replicate-strategy decisions, which differs from the conditional-mean expiry-support decision considered here.
Throughout, the primary cutoff is 10\%.
The 10\% threshold is not intended as an optimal inferential cutoff.
It is a pre-specified operational rule for identifying fitted random-effect contributions that are small enough to be practically negligible at the proposed expiry.
Because variance-component decisions at zero are non-regular, any fixed threshold is necessarily heuristic; we therefore evaluate 5\%, 10\%, and 20\% thresholds in Appendix~\ref{app:additional_sensitivity}. In those analyses, the 5\% and 10\% thresholds have similar qualitative behavior, whereas 20\% increases support probability but also produces more undercoverage in the small-to-moderate variance region. This motivates the more conservative 10\% rule for this expiry-support setting. The $p$-scale calculation in Section~\ref{sec:vcond_mechanism} gives a design-based interpretation of this range: in the balanced random-intercept calculation, the half-saturation point is $p^*=1/(m+1)$, so 10\% is the transition point for the 9-pull worked example. For the simulation design with $m=7$ scheduled time points, the same balanced calculation gives $p^*=1/(m+1)=0.125$, so the 10\% operational threshold lies just below the simulation-design half-saturation point. The 5\%, 10\%, and 20\% sensitivity analyses span this neighborhood. Thus the rule remains heuristic, but it is positioned near the transition from the steep near-boundary part of the prediction-variance curve to the flatter region.

\paragraph{Random-intercept fit.}
After fitting Model~(2), compute the estimated lot variance fraction
\[
\widehat{\mathrm{vcfrac}}=\frac{\widehat\sigma_{b0}^2}{\widehat\sigma_{b0}^2+\widehat\sigma_e^2}.
\]
If $\widehat{\mathrm{vcfrac}}<0.10$, refit as pooled OLS (Model~(3)).

\paragraph{Random intercept-and-slope fit.}
For Model~(1), let $t_*$ denote the proposed expiry month (here $t_*=48$ months in the simulation study).
With uncorrelated random intercept and slope, the marginal variance of $Y$ at time $t_*$ is $\sigma_{b0}^2+t_*^2\sigma_{b1}^2+\sigma_e^2$.
Define the estimated intercept- and slope-variance contributions at $t_*$ as
\[
\widehat p_{b0}(t_*)=\frac{\widehat\sigma_{b0}^2}{\widehat\sigma_{b0}^2+t_*^2\widehat\sigma_{b1}^2+\widehat\sigma_e^2},
\qquad
\widehat p_{b1}(t_*)=\frac{t_*^2\widehat\sigma_{b1}^2}{\widehat\sigma_{b0}^2+t_*^2\widehat\sigma_{b1}^2+\widehat\sigma_e^2}.
\]
These are the full Model~(1) quantities reported in Table~\ref{tab:vc_zero_freq}.
The reduction rule first checks the slope contribution: if $\widehat p_{b1}(t_*)<0.10$, drop the random slope and refit Model~(2).
In the Model~(2) refit, compute $\widehat{\mathrm{vcfrac}}$ as in Eq.~\eqref{eq:vcfrac}; if $\widehat{\mathrm{vcfrac}}<0.10$, drop the random intercept and refit as pooled OLS (Model~(3)).

\subsection{AICc step-down procedure and boundary non-regularity}
\label{sec:aicc}

Both the variance-contribution reduction rule (Section~\ref{sec:reduced}) and the 
information-criterion step-down described below implicitly decide whether a variance 
component is negligible. Variance components are constrained to be nonnegative and the 
null value 0 lies on the boundary of the parameter space, so such decisions are 
non-regular.\citep{selfliang,Scheipl2008} (See Section~\ref{sec:boundary_modelsel} 
for further discussion.) Nevertheless, information-criterion step-down procedures have been used in practice for stability model selection \citep{Pack2018Stability}, so we include AICc as a pragmatic comparator and evaluate its empirical behavior in our near-boundary expiry settings.

The AICc procedure fits candidate Models~(1)--(3) and selects the model with the 
smallest AICc.\citep{Sugiura1978,HurvichTsai1989,Pack2018Stability}
Because all candidate models share the same fixed-effects specification (intercept and slope), REML-based AICc values are comparable across these models.
In the simulations, mixed-model AICc values were taken from the \texttt{PROC MIXED} fit-statistics table for each candidate REML fit.
Conditional-mean confidence limits are then computed from the selected model using SAT when a mixed model is selected (Model~(1) or~(2)), and using residual degrees of freedom when pooled OLS is selected (Model~(3)).
Because AICc selection depends on the likelihood and is performed before the DDF method is applied, SAT\_AICc and CONTAIN\_AICc select the same model in each dataset; in our simulations their operating characteristics were nearly indistinguishable, so for brevity we present SAT\_AICc only.

\paragraph{Additional COVTEST reduction comparator.}
As an additional appendix sensitivity analysis, we also evaluate a p-value-based random-effect reduction procedure inspired by the fixed-effect pooling cutoffs in Q1E.
This procedure, denoted SAT\_covtest25, uses raw \texttt{PROC MIXED} \texttt{COVTEST} output in the same conditional Model~(1) $\rightarrow$ Model~(2) $\rightarrow$ Model~(3) step-down sequence as the variance-contribution reduction rule. It drops a random-effect term when the corresponding variance-component test has $p>0.25$, and then computes conditional-mean limits using SAT or OLS as applicable.
We treat this p-value procedure as a sensitivity analysis only, because tests of variance components at zero are non-regular and standard chi-squared reference distributions do not provide ordinary interior-parameter inference.\citep{selfliang,Scheipl2008}

\subsection{Fixed-lot (Q1E-style) poolability comparator}
\label{sec:fixedq1e}

We include a Q1E-style fixed-lot ANCOVA comparator in which lot is treated as fixed and pooling follows the step-down logic in ICH Q1E using a $p=0.25$ cutoff.\citep{ICHQ1E2003}
We fit a fixed-effects model with lot and lot-by-time terms, test slope poolability, and if slopes are pooled, test intercept poolability; when pooling is accepted, the model reduces to pooled regression.
This step-down procedure corresponds to the fixed-lot analogs of Models~(1)--(3). We include it to anchor the mixed-model comparisons to the current Q1E-style fixed-lot practice, not because it targets the same random-lot conditional-mean estimand as the mixed-model methods.

\subsection{Software implementation}
\label{sec:software}

All mixed-model simulation fits were performed in SAS \texttt{PROC MIXED} using restricted maximum likelihood estimation with bounded variance components (nonnegative constraints; the default in \texttt{PROC MIXED}).
This constraint is required because the conditional-mean $t$-based pointwise limits used in the profiler displays and decision summaries are not produced when a fitted variance component is negative.
For Model~(1) fits, the random intercept and random slope were specified with separate \texttt{RANDOM} statements and \texttt{TYPE=VC} to enforce an uncorrelated random-effects structure (covariance fixed at zero).
For the worked example, JMP was used only for cross-software illustration. To make conditional-mean limits defined and comparable across platforms, the JMP variance-component setting was changed from the default unbounded setting to bounded. Base JMP users should similarly use bounded variance components when conditional-mean limits are required. Both figures use bounded variance components; Figure~\ref{fig:jmp-default} additionally retains JMP's default Satterthwaite DDF, while Figure~\ref{fig:jmp-contain} uses containment.
Simulation code, simulation outputs, and the worked example dataset are available in the data repository.\citep{KarlMendeleyData2025} Additional implementation details for JMP are available elsewhere.\citep{Pack2018Stability}

For the unreduced SAT and CONTAIN analyses, zero variance-component estimates were treated as boundary estimates from the specified model, not as automatic model-selection events. Thus, random-effect terms were retained unless a reduced procedure explicitly removed them and refit the model. A zero variance component contributes no estimated variability from that term, so estimates, standard errors, and conditional-mean intervals may coincide with those from the corresponding reduced model. The degrees-of-freedom consequences, however, depend on the DDF method: SAT can behave like the reduced covariance model at exact zero, whereas CONTAIN degrees of freedom can depend on the retained random-effect structure.

For the reduced-variant procedure (Section~\ref{sec:reduced}), the full model was fit, the relevant estimated variance contribution(s) were computed, and a simplified model was refit when the threshold rule triggered.

Conditional-mean confidence limits were extracted from the \texttt{OUTP=} dataset in \texttt{PROC MIXED}, which reports predicted values, their standard errors, denominator degrees of freedom, and $t$-based confidence limits.\citep{SASMixedDDFM152}
In all simulations, \texttt{ALPHAP=0.10} was used so that the reported two-sided 90\% limits correspond to a one-sided 95\% lower confidence limit for each conditional mean.
The key per-row variables are \texttt{Pred} (predicted conditional mean $\widehat{\mu}_i(t)$), \texttt{StdErrPred} (standard error $\widehat{\mathrm{se}}\{\widehat{\mu}_i(t)\}$), \texttt{DF} (denominator degrees of freedom $\nu(t,i)$), and \texttt{Lower}/\texttt{Upper} ($t$-based confidence limits for the conditional mean determined by \texttt{ALPHAP}).

\subsection{Simulation design}
\label{sec:simdesign}

We conduct a simulation study to quantify the impact of near-boundary variance-component estimates on conditional-mean inference and on an extrapolated proposed-expiry decision at $t_*=48$ months.

\paragraph{Data-generating process.}
Simulated datasets are generated under the random-intercept model (Model~(2)) with $L=10$ lots and a balanced schedule $t\in\{0,3,6,9,12,24,36\}$ months (7 observations per lot; $n=70$).
Fixed effects are $\beta_0=100$ and $\beta_1=-10/57$, so the population mean crosses the specification limit ($\mathrm{LSL}=90$) at 57 months.

We fix the total variance at $\sigma_{\mathrm{tot}}^2=1$ and vary the true lot-intercept variance fraction
\begin{equation}
\mathrm{vcfrac}_{\mathrm{true}}=\frac{\sigma_{b0}^2}{\sigma_{b0}^2+\sigma_e^2},
\label{eq:vcfract}
\end{equation}
so that $\sigma_{b0}^2=\mathrm{vcfrac}_{\mathrm{true}}$ and $\sigma_e^2=1-\mathrm{vcfrac}_{\mathrm{true}}$.
The data-generating process contains no random-slope heterogeneity ($\sigma_{b1}^2=0$).

\paragraph{Decision criterion at $t_*=48$ months.}
For each fitted dataset, we compute the lot-specific conditional-mean lower confidence limits at 48 months and record whether
\[
\min_i \mathrm{LCL}_i(48)\ge \mathrm{LSL}.
\]
We report the proportion of simulated datasets meeting this criterion, denoted $\Pr(\mathrm{Support48})$.

\paragraph{Analysis models and inference approaches.}
Each dataset is analyzed using the methods in Table~\ref{tab:approaches}.
For the decision operating-characteristic curves (Section~\ref{sec:sim_support}) and the coverage diagnostic (Section~\ref{sec:sim_coverage}), mixed-model analyses begin with the full random-intercept-and-slope fit (Model~(1)) with uncorrelated random effects, even though the data-generating process has $\sigma_{b1}^2=0$.
This choice is not meant to assume that Model~(2) is universally the most likely stability data-generating mechanism.
Rather, it isolates a common operational setting in which slope heterogeneity is scientifically plausible enough to include in the initial model, but the true or fitted random-slope variance may be small.
The mechanism plots in Section~\ref{sec:sim_width_df} fit the generating Model~(2) directly, showing that the SAT collapse is not solely an artifact of random-slope overparameterization.
The SAT\_reduced and SAT\_AICc procedures may step down by dropping the slope and/or intercept, potentially selecting Model~(2) or pooled OLS (Model~(3)).
To isolate the random-intercept boundary behavior without the additional slope term, the interval-margin and DDF summaries in Section~\ref{sec:sim_width_df} fit Model~(2).

\paragraph{Simulation grid and reference curves.}
We repeat the simulation across a grid of $\mathrm{vcfrac}_{\mathrm{true}}$ values spanning $[0,1)$. The main tabular grid is $0,0.1,\ldots,0.9$; several decision and sensitivity curves include additional near-one grid values below 1 to show right-tail behavior.
Monte Carlo sizes vary by diagnostic (200 datasets per setting for margin/DDF summaries, 500 for Table~\ref{tab:vc_zero_freq}, 1500 for the coverage diagnostic, and 2000 for the decision curves); the relevant sample sizes are stated in the corresponding figure captions and table notes.
For decision curves, we also plot an ``Expected'' reference curve computed under the data-generating random-intercept model with known variance components (Appendix~\ref{app:baselinepass_lsl}). This known-variance reference omits variance-component estimation, boundary truncation, DDF approximation, and data-dependent model-selection effects; it is a reference calculation rather than a calibration target for procedures that estimate or select variance components. While the main simulation reference curve does not include random-slope variability, the matched Model~(1) sensitivity in Appendix~\ref{app:m1_sensitivity} uses the corresponding random-intercept-and-slope known-variance reference.

\paragraph{Additional sensitivity analyses.}
To address model-matching and reduction-rule questions,
we also evaluate matched random-intercept-and-slope Model~(1) simulations with low and moderate random-slope variance contributions at $t_*=48$. The Model~(3)-true case is already included in the main simulation grid at $\mathrm{vcfrac}_{\mathrm{true}}=0$, where $\sigma^2_{b0}=\sigma^2_{b1}=0$.
For the Model~(1) simulations, we keep $\sigma_{b0}^2+\sigma_e^2=1$ and choose $\sigma_{b1}^2$ so that the random-slope contribution to the marginal variance at $t_*=48$ months is either 5\% or 12.5\%; this gives $\sigma_{b1}^2=2.28\times 10^{-5}$ and $6.20\times 10^{-5}$, respectively.
Appendix~\ref{app:additional_sensitivity} also reports sensitivity analyses for Kenward--Roger, 5/10/20\% threshold sensitivity, and SAT\_covtest25.
\subsection{Worked example dataset}
\label{sec:workedexample}

We use a representative simulated dataset to illustrate how DDF choices affect conditional-mean profiler output in JMP and SAS. The worked example comprises 14 lots with 9 scheduled pull times through 60 months (0, 3, 6, 9, 12, 24, 36, 48, 60). The simulation study uses a different balanced design (10 lots, 7 pull times through 36 months; Section~\ref{sec:simdesign}). We intentionally use a larger worked example, with more lots, more observations per lot, and an approximately flat mean trend, so that the profiler overlay focuses on the DDF mechanism rather than the degradation slope. This also shows that the boundary-proximal SAT DDF collapse is not a small-sample artifact. The example was generated without lot-to-lot variability ($\mathrm{vcfrac}_{\mathrm{true}}=0$) but selected to yield a small positive bounded-REML lot variance estimate, where differences among DDF methods are most apparent. The worked example dataset is available in the data repository.\citep{KarlMendeleyData2025}

\section{Results}

\subsection{Worked example: conditional profilers and software defaults}

\begin{figure}[htbp]
\centering
\includegraphics[width=0.75\linewidth]{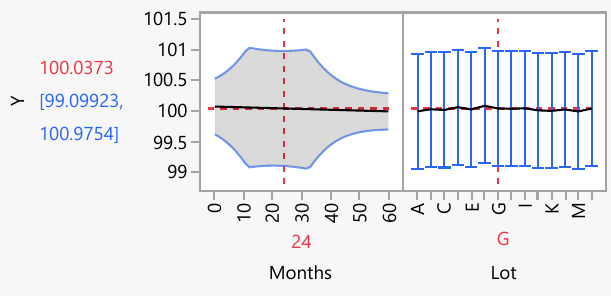}
\caption{Worked example. JMP Conditional Model Profiler for Lot~G using JMP's default Satterthwaite DDF method for conditional-mean display limits. Display intervals are two-sided 95\%.}
\label{fig:jmp-default}
\end{figure}

\begin{figure}[htbp]
\centering
\includegraphics[width=0.75\linewidth]{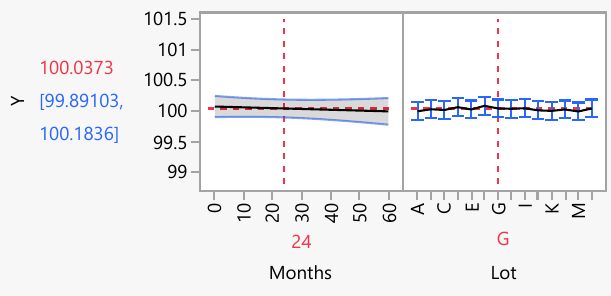}
\caption{Worked example. JMP Conditional Model Profiler for the same lot after switching the DDF method to containment. Display intervals are two-sided 95\%.}
\label{fig:jmp-contain}
\end{figure}

Figures~\ref{fig:jmp-default} and \ref{fig:jmp-contain} show JMP Conditional Model Profiler output for Lot~G under two inference settings.
By default, JMP uses a Satterthwaite denominator degrees-of-freedom approximation for conditional-mean confidence limits \citep{JMPKH}; in JMP~19 Pro it is also possible to request containment DDF for these intervals, which we use for Figure~\ref{fig:jmp-contain}.
With the default Satterthwaite DDF method and bounded variance components, the conditional-mean confidence limits widen at intermediate months and narrow again at later months, producing a nonmonotone sequence of pointwise limits on the scheduled grid.
After switching the DDF method to containment (with bounded variance components), the conditional-mean limits do not exhibit the same nonmonotone widening (Figure~\ref{fig:jmp-contain}).
The analytical mechanism behind this behavior is given in Section~\ref{sec:vcond_mechanism} and Appendix~\ref{app:vcond_derivation}.

\begin{figure}[htbp]
\centering
\includegraphics[width=0.75\linewidth]{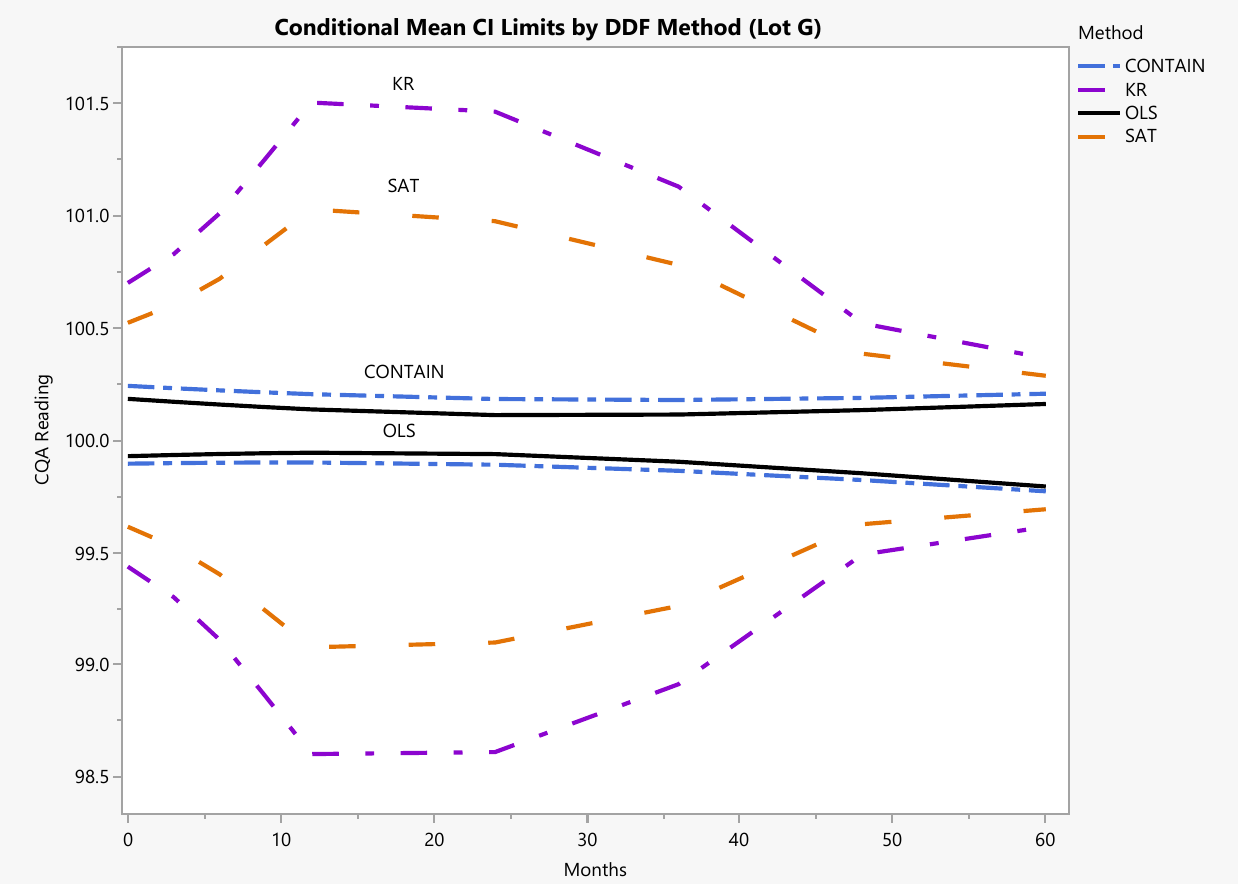}
\caption{Worked example. Conditional-mean display limits by DDF method for Lot~G from \texttt{PROC MIXED}. Display intervals are two-sided 95\%. OLS provides the pooled regression reference for this simulated example, which was generated with $\mathrm{vcfrac}_{\mathrm{true}}=0$; KR denotes \texttt{DDFM=KR2}.}
\label{fig:sas-overlay}
\end{figure}

Figure~\ref{fig:sas-overlay} provides the analogous display in \texttt{PROC MIXED}, overlaying conditional-mean confidence limits for the same lot under OLS, containment, Satterthwaite, and Kenward--Roger. For visual comparability, the worked-example displays in Figures~\ref{fig:jmp-default}--\ref{fig:sas-overlay} use two-sided 95\% pointwise display intervals to make the DDF-driven symmetric widening and nonmonotonicity visually clear. The formal expiry-decision analyses and numerical Appendix~\ref{app:satt_rebuild} calculations use the one-sided 95\% lower-limit convention in Eq.~\eqref{eq:lcl}, implemented in \texttt{PROC MIXED} as two-sided 90\% prediction limits with \texttt{ALPHAP=0.10}.

For comparability with the mechanism plots in Section~\ref{sec:sim_width_df}, the \texttt{PROC MIXED} overlays are based on the random-intercept model (Model~(2)), so the behavior shown here  is not caused by random-slope overparameterization.
Among the mixed-model fits shown here, the fitted conditional mean is unchanged by the DDF method; the pooled OLS reference is also similar for this simulated example.
The difference is in the confidence limits: Satterthwaite and KR produce markedly wider and nonmonotone limits when the fitted lot variance is small but positive.
Because this simulated worked example was generated with $\mathrm{vcfrac}_{\mathrm{true}}=0$, pooled OLS corresponds to the data-generating model.
Under bounded REML, the fitted lot variance fraction from the Model~(2) fit is $\widehat{\mathrm{vcfrac}} \approx 0.019$ (1.9\%), placing it in the small-variance setting where variance-component-dependent DDF collapse is most severe.
Specification limits are omitted to emphasize the relative shape and width of the confidence limits across methods.

For the random-intercept mixed-model calculation underlying the Satterthwaite and containment intervals in this display, the prediction time $t_*$ enters the conditional prediction variance only through the fixed-effect leverage $A_*=x_*^\top(X^\top X)^{-1}x_*$ (Section~\ref{sec:vcond_mechanism} and Appendix~\ref{app:vcond_derivation}).
Thus the corresponding prediction standard error is smallest near the time-design centroid; the pronounced mid-design widening of the variance-component-dependent confidence limits in Figure~\ref{fig:sas-overlay} is primarily a collapsed-DDF $t$-multiplier effect, not evidence of weaker data support at those months.
At the same Lot~G, Month~24 conditional scoring point, \texttt{PROC MIXED} KR output reports DDF=1, the same one-sided $t$ multiplier as SAT, and a larger adjusted prediction standard error, which makes the KR interval wider than the SAT interval (Appendix~\ref{app:satt_rebuild}).

\subsection{Simulation results: interval margin and denominator degrees of freedom}
\label{sec:sim_width_df}

To show that the DDF collapse can arise from boundary-proximal variance components even without random-slope overparameterization, this subsection analyzes each simulated dataset using the random-intercept model (Model~(2)), so that the only random-effect variance component that can approach the boundary is the lot-intercept variance.
(All datasets are generated under Model~(2); Section~\ref{sec:simdesign}.)
We summarize inference behavior using the lower 95\% conditional-mean margin,
defined pointwise as
\[
w_i(t)=\widehat\mu_i(t)-\mathrm{LCL}_i(t)
      = t_{0.95,\nu(t,i)}\,\widehat{\mathrm{se}}\{\widehat\mu_i(t)\}.
\]
In \texttt{PROC MIXED} this is obtained from \texttt{OUTP} by requesting two-sided 90\%
limits for the predicted conditional mean (\texttt{ALPHAP}=0.10) and taking $w_i(t)=\texttt{Pred}-\texttt{Lower}$.

We index results in two ways: by the fitted variance fraction $\widehat{\mathrm{vcfrac}}$ (Eq.~\eqref{eq:vcfrac} with estimated variance components) and by the true generating fraction $\mathrm{vcfrac}_{\mathrm{true}}$ (Section~\ref{sec:simdesign}).
Plotting against $\widehat{\mathrm{vcfrac}}$ reveals the mechanism: the SAT DDF collapse and margin inflation are driven by the realized fitted variance component, and the relationship is most clearly visible when conditioning on the estimated quantity.
Plotting against $\mathrm{vcfrac}_{\mathrm{true}}$ addresses the practical question of expected behavior for a given attribute, which in practice has a fixed but unknown lot-to-lot variance; each true-variance setting produces a mixture of fitted values (some at the boundary, some small-positive, some larger), so the true-variance plots show the marginal behavior induced by this mixture.

\begin{figure}[htbp]
\centering
\begin{subfigure}[t]{0.95\textwidth}
\centering
\includegraphics[width=\linewidth]{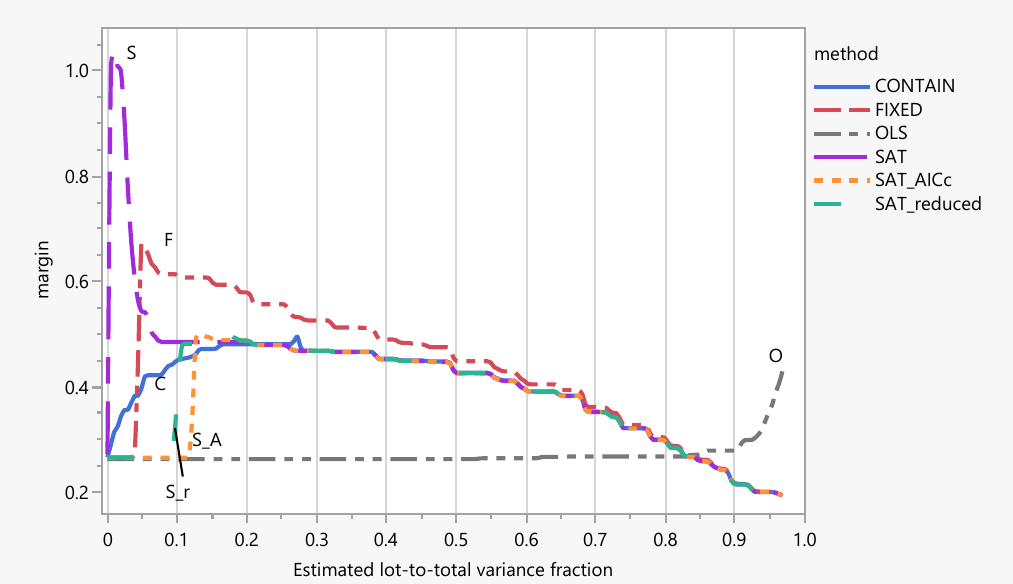}
\caption{Mean margin versus the fitted $\widehat{\mathrm{vcfrac}}$.}
\label{fig:sim_width_est}
\end{subfigure}

\vspace{0.8em}

\begin{subfigure}[t]{0.95\textwidth}
\centering
\includegraphics[width=\linewidth]{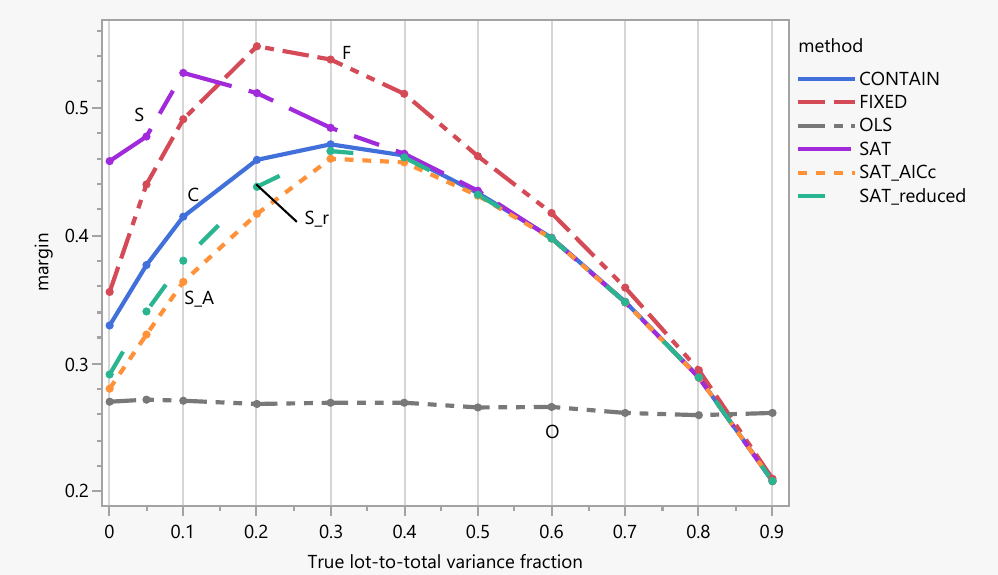}
\caption{Mean margin versus the true $\mathrm{vcfrac}_{\mathrm{true}}$.}
\label{fig:sim_width_true}
\end{subfigure}
\caption{Mean lower 95\% mean-limit margin for each implemented procedure, averaged over lots and scheduled months within each dataset. For the random-lot mixed-model methods, this is the conditional-mean LCL margin. Inline letters identify the same methods as the legend (S=SAT, C=CONTAIN, F=FIXED, O=OLS, S\_A=SAT\_AICc, S\_r=SAT\_reduced). Each grid value of $\mathrm{vcfrac}_{\mathrm{true}}$ uses 200 simulated datasets; all methods are applied to the same datasets at each grid value.}
\label{fig:sim_width}
\end{figure}

\begin{figure}[htbp]
\centering
\begin{subfigure}[t]{0.95\textwidth}
\centering
\includegraphics[width=\linewidth]{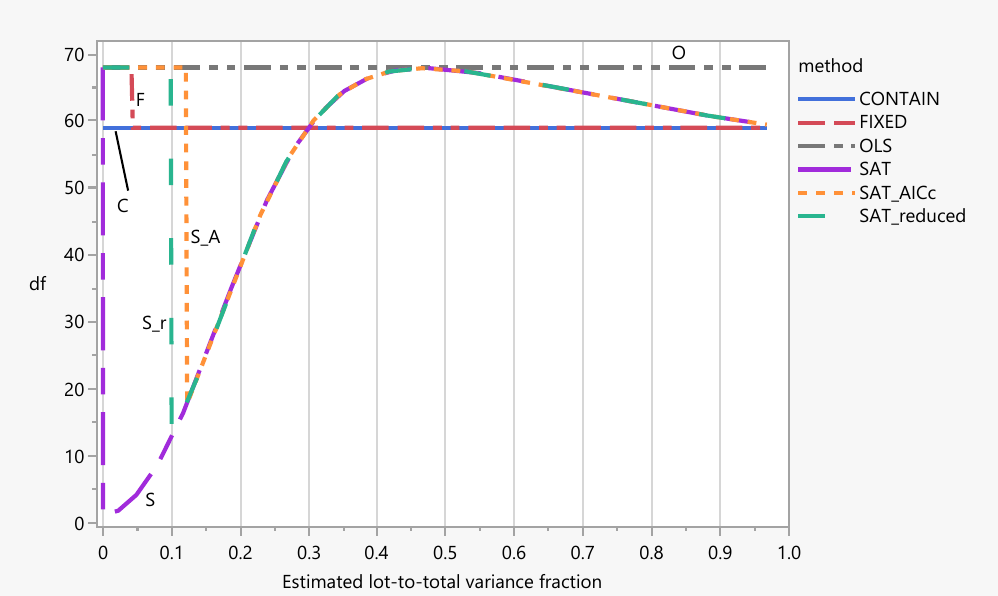}
\caption{Mean DDF versus the fitted $\widehat{\mathrm{vcfrac}}$.}
\label{fig:sim_df_est}
\end{subfigure}

\vspace{0.8em}

\begin{subfigure}[t]{0.95\textwidth}
\centering
\includegraphics[width=\linewidth]{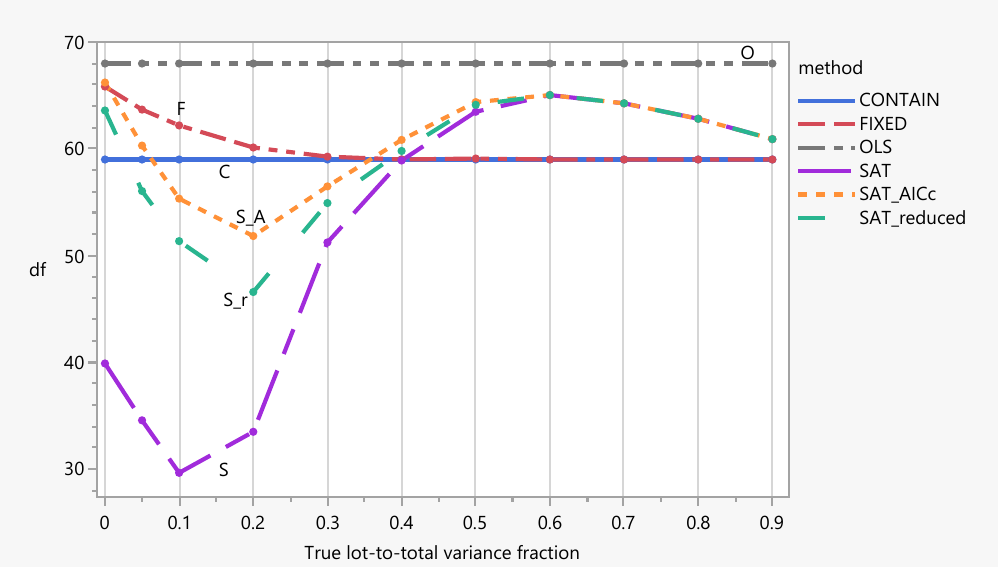}
\caption{Mean DDF versus the true $\mathrm{vcfrac}_{\mathrm{true}}$.}
\label{fig:sim_df_true}
\end{subfigure}
\caption{Mean denominator degrees of freedom (DDF) associated with the implemented mean limits, averaged over lots and scheduled months within each dataset. For the random-lot mixed-model methods, these are the DDF for conditional-mean confidence limits. Inline letters identify the same methods as the legend (S=SAT, C=CONTAIN, F=FIXED, O=OLS, S\_A=SAT\_AICc, S\_r=SAT\_reduced). Each grid value of $\mathrm{vcfrac}_{\mathrm{true}}$ uses 200 simulated datasets; all methods are applied to the same datasets at each grid value.}
\label{fig:sim_df}
\end{figure}

Figures~\ref{fig:sim_width} and~\ref{fig:sim_df} plot the mean lower-limit margin and mean DDF, each averaged over lots and scheduled months within a dataset and then averaged over 200 datasets per setting.
Several patterns are relevant.
First, Satterthwaite inference changes sharply as $\widehat{\mathrm{vcfrac}}$ approaches zero from above: SAT DDF collapses (Figure~\ref{fig:sim_df_est}), inflating the $t$ critical value and producing much larger lower confidence-limit margins (and therefore lower LCLs) than either containment or OLS (Figure~\ref{fig:sim_width_est}).
At the exact-zero boundary, SAT coincides with the reduced or OLS covariance model in this one-component setting because a zero covariance parameter contributes no estimated variance-component uncertainty to the SAT DDF calculation.
This is a DDF-calculation effect, not an automatic refit: the random-effect term remains in the unreduced model specification.
By contrast, containment continues to use the retained random-effect design in its DDF assignment; in this balanced design, containment DDF equals $n-\mathrm{rank}([X\ Z])=70-11=59$, reflecting the collinearity between the fixed-effect intercept column and the lot-indicator columns in $Z$ \citep{SASMixedDDFM152}.
Thus, CONTAIN does not jump to the OLS DDF at the boundary and avoids the sharp SAT discontinuity.
Immediately to the right of the boundary, however, small positive variance estimates give much smaller SAT DDF and wider confidence limits.
Second, SAT\_reduced and SAT\_AICc mitigate extreme SAT behavior by stepping down to simpler models when fitted random-effect contributions at the proposed expiry are judged negligible (10\% cutoff for SAT\_reduced, or AICc selection for SAT\_AICc); their curves therefore exhibit discrete changes near the model-switching region.

The preceding results fit the generating model (Model~(2)) to isolate the DDF mechanism.
The decision and coverage analyses then use the full Model~(1) starting fit described in Section~\ref{sec:simdesign}, so that the same near-boundary mechanism is evaluated in the overparameterized workflow encountered in routine use.
Additional Kenward--Roger mechanism simulations in Appendix~\ref{app:additional_sensitivity} show the same near-boundary DDF collapse as SAT; at small positive fitted variance-fraction values, KR margins are even larger because the KR-adjusted prediction standard errors are larger.

\subsection{Frequency of boundary variance-component estimates under the full random-intercept-and-slope fit}
\label{sec:vc_zero}

For the decision and coverage simulations (Sections~\ref{sec:sim_coverage} and~\ref{sec:sim_support}), mixed-model analyses begin with the full random-intercept-and-slope fit (Model~(1)) under bounded REML, even though the data-generating process has no random slope ($\sigma_{b1}^2=0$).
In this overparameterized setting, fitted variance components can be estimated at the boundary ($\hat\sigma^2=0$) with non-negligible frequency, and even very small positive slope-variance estimates can contribute appreciably to the marginal variance at late time points through the $t_*^2$ scaling.

Table~\ref{tab:vc_zero_freq} summarizes boundary frequencies and fitted variance contributions at $t_*=48$ months. These boundary frequencies indicate estimates from the full bounded-REML fit, not automatic removal of the corresponding random-effect terms.
The random-slope variance is estimated at the boundary in about 0.52--0.57 of datasets across the grid.
Across all datasets, the mean fitted slope contribution to the marginal variance at 48 months is amplified by $t_*^2$, with $\overline{\hat p}_{b1}(48)$ ranging from about 0.05 to 0.16 across settings.
These near-boundary fits correspond to the variance-component range in which variance-component-dependent DDF methods such as SAT can become unstable, and in which reduction or information-criterion step-down rules most often alter the selected model.

\begin{table}[htbp]
\centering
\caption{Boundary frequencies and mean fitted variance contributions at $t_*=48$ months under the full random-intercept-and-slope fit (Model~(1)). The final two columns report $\overline{\hat p}_{b0}(48)$ and $\overline{\hat p}_{b1}(48)$, the mean estimated intercept and slope contributions to the marginal variance at $t_*=48$ as defined in Section~\ref{sec:reduced}. Each grid value of $\mathrm{vcfrac}_{\mathrm{true}}$ uses 500 simulated datasets.}
\label{tab:vc_zero_freq}
\small
\begin{tabular}{rcccc}
\toprule
$\mathrm{vcfrac}_{\mathrm{true}}$ & $\Pr(\hat\sigma_{b0}^2=0)$ & $\Pr(\hat\sigma_{b1}^2=0)$ & $\overline{\hat p}_{b0}(48)$ & $\overline{\hat p}_{b1}(48)$ \\
\midrule
0.0 & 0.618 & 0.570 & 0.022 & 0.129 \\
0.1 & 0.246 & 0.558 & 0.078 & 0.146 \\
0.2 & 0.080 & 0.542 & 0.163 & 0.157 \\
0.3 & 0.046 & 0.568 & 0.231 & 0.148 \\
0.4 & 0.012 & 0.536 & 0.322 & 0.137 \\
0.5 & 0.000 & 0.522 & 0.406 & 0.136 \\
0.6 & 0.002 & 0.524 & 0.504 & 0.129 \\
0.7 & 0.000 & 0.534 & 0.613 & 0.095 \\
0.8 & 0.002 & 0.546 & 0.705 & 0.082 \\
0.9 & 0.000 & 0.562 & 0.836 & 0.049 \\
\bottomrule
\end{tabular}
\end{table}
\subsection{Coverage diagnostic for conditional-mean confidence limits at 48 months}
\label{sec:sim_coverage}

As a diagnostic separate from the expiry decision rule, we evaluate the empirical pointwise coverage of the nominal 95\% one-sided lower confidence limits for the lot-specific conditional means at $t_*=48$ months.
Because these limits are EBLUP-based, this coverage calculation is an operating-characteristic diagnostic for the implemented pointwise procedure, not a claim of exact finite-sample coverage for realized random effects.
BLUPs are shrinkage predictors of realized lot effects rather than fixed-effect estimates of lot-specific parameters,\citep{robinson1991blup} so nominal pointwise coverage is a useful benchmark rather than a guaranteed finite-sample target.
For each simulated dataset and lot, we compute whether the lower confidence limit covers the true conditional mean, i.e.,
\[
\mathrm{LCL}_i(48)\le \mu_i(48),
\]
where $\mu_i(48)$ is the true conditional mean under the data-generating model.
We report the proportion of lots meeting this condition, pooled across all lot--dataset combinations.

\begin{figure}[htbp]
\centering
\includegraphics[width=0.8\linewidth]{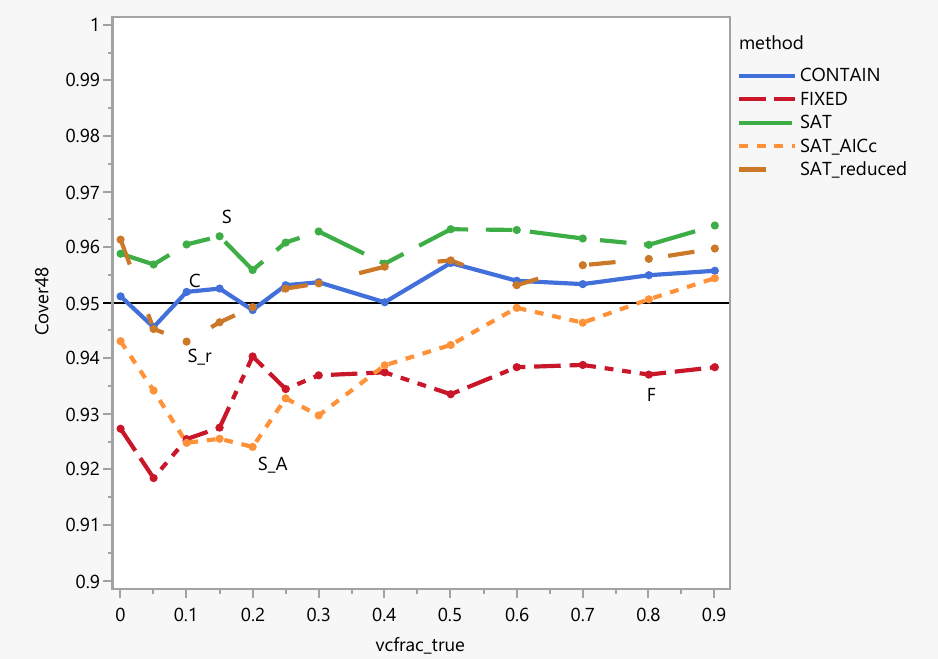}
\caption{Empirical coverage at $t_*=48$ months of nominal 95\% one-sided lower confidence limits for $\mu_i(48)$, computed as $\Pr\{\mathrm{LCL}_i(48)\le \mu_i(48)\}$, pooled across all lot--dataset combinations. Each grid value of $\mathrm{vcfrac}_{\mathrm{true}}$ uses 1500 simulated datasets; all methods are applied to the same datasets at each grid value.}
\label{fig:coverage_tp48}
\end{figure}

Figure~\ref{fig:coverage_tp48} shows that CONTAIN stays close to the nominal 0.95 target across settings (0.946--0.957), while SAT is slightly conservative (0.955--0.964). This slight overcoverage is the coverage-probability consequence of the inflated SAT lower-limit margins documented in Section~\ref{sec:sim_width_df}: lower confidence limits that are farther below the fitted mean are less likely to exceed the true conditional mean, but they also reduce power (Section~\ref{sec:sim_support}).
The fixed-lot comparator undercovers (0.918--0.940), reflecting that it does not target a random-lot conditional mean and does not borrow information across lots in the same way.
This provides useful context for the sensitivity analyses below: modest departures from nominal coverage for mixed-model procedures should be interpreted relative to both the nominal 0.95 benchmark and the Q1E-style fixed-lot comparator.
SAT\_reduced closely tracks SAT, indicating that the reduction rule primarily affects the most extreme near-boundary cases rather than the average coverage.
SAT\_AICc shows modest undercoverage in the small-to-moderate $\mathrm{vcfrac}_{\mathrm{true}}$ setting (minimum 0.924 at $\mathrm{vcfrac}_{\mathrm{true}}=0.20$).
Appendix~\ref{app:threshold_sensitivity}, Appendix~\ref{app:covtest_sensitivity}, and Appendix~\ref{app:m1_sensitivity} provide sensitivity analyses for threshold choice, the SAT\_covtest25 p-value reduction comparator, and matched Model~(1) data generation, respectively.
In those analyses, the 5\% and 10\% SAT\_reduced thresholds remain close to nominal coverage, while the 20\% threshold can fall below nominal coverage in the small-to-moderate variance region (minimum 0.927).
The SAT\_covtest25 comparator similarly undercovers in the low-variance region (minimum 0.934, compared with 0.947 for SAT\_reduced), supporting its use as a sensitivity analysis rather than a primary procedure.
When the data-generating mechanism includes a nonzero random slope, pointwise coverage becomes a more challenging operating-characteristic diagnostic for EBLUP-based lot-specific predictions: SAT is closest to nominal among the mixed-model procedures (0.936--0.953), whereas CONTAIN is modestly below nominal in the moderate-slope/high-intercept-variance corner (minimum 0.911; Appendix~\ref{app:additional_sensitivity}).

\subsection{Decision operating characteristics at the proposed expiry time}
\label{sec:sim_support}

\begin{figure}[htbp]
\centering
\includegraphics[width=0.85\linewidth]{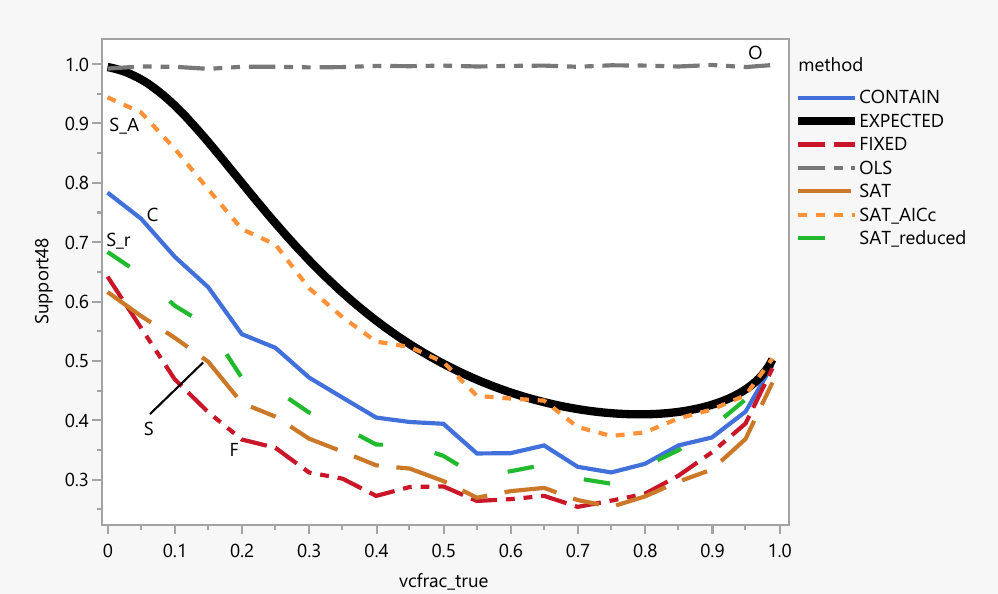}
\caption{Decision operating characteristic curves for $\Pr(\mathrm{Support48})$ at $t_*=48$ months under the baseline calibration (population mean crosses the LSL at 57 months). ``Expected'' denotes the known-variance reference curve from Appendix~\ref{app:baselinepass_lsl}. Each grid value of $\mathrm{vcfrac}_{\mathrm{true}}$ uses 2000 simulated datasets; all methods are applied to the same datasets at each grid value.}
\label{fig:pass48_prob}
\end{figure}

In a shelf-life decision setting, the primary operating characteristic is $\Pr(\mathrm{Support48})$, the probability that the minimum conditional-mean lower confidence limit across lots at 48 months is at or above the LSL.
Figure~\ref{fig:pass48_prob} shows $\Pr(\mathrm{Support48})$ as a function of $\mathrm{vcfrac}_{\mathrm{true}}$ under the baseline calibration (population mean crossing at 57 months), along with an ``Expected'' known-variance reference curve computed under known variance components (Appendix~\ref{app:baselinepass_lsl}).

Several features are prominent.
First, pooled OLS is optimistic across the grid (about 0.99), because it ignores between-lot variability. We include OLS to capture the data-generating model when the true between-lot variance is zero ($\mathrm{vcfrac}_{\mathrm{true}}=0$).
Second, when $\mathrm{vcfrac}_{\mathrm{true}}$ is small, the full mixed-model analyses are conservative relative to the known-variance reference curve, and the choice of DDF method matters.
For example, at $\mathrm{vcfrac}_{\mathrm{true}}=0$ the known-variance reference is 0.995, while CONTAIN and SAT yield 0.784 and 0.616, respectively; the reduction rule increases SAT to 0.684.
Third, the AICc step-down tracks the reference curve closely in the mid-range where the random-intercept component is identifiable; at $\mathrm{vcfrac}_{\mathrm{true}}=0.50$, SAT\_AICc yields 0.498 versus 0.495 for the reference curve.

The fixed-lot comparator is consistently conservative in this random-lot setting (e.g., 0.289 at $\mathrm{vcfrac}_{\mathrm{true}}=0.50$), reflecting the lack of shrinkage and the effective requirement that each estimated lot effect be sufficiently above the LSL.
CONTAIN provides a substantial power gain relative to FIXED; for example, at $\mathrm{vcfrac}_{\mathrm{true}}=0.50$ CONTAIN supports the proposed expiry in 0.394 of datasets (known-variance reference 0.495).
The appendix sensitivity analyses give the same qualitative trade-off: more aggressive reduction or pooling can increase Support48, but this gain can coincide with lower pointwise coverage or greater sensitivity when true random-effect variation is small but nonzero.

\begin{figure}[htbp]
\centering
\includegraphics[width=0.85\linewidth]{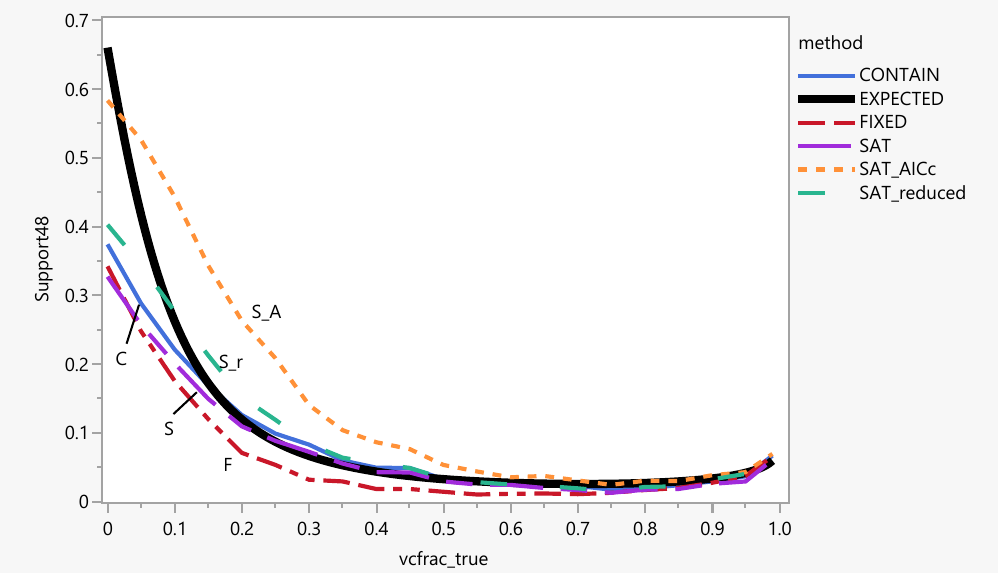}
\caption{Sensitivity analysis: decision operating characteristic curves for $\Pr(\mathrm{Support48})$ at $t_*=48$ months under a tighter mean trend (population mean crosses the LSL at 52 months). ``Expected'' denotes the known-variance reference curve under the corresponding mean trend. Each grid value of $\mathrm{vcfrac}_{\mathrm{true}}$ uses 2000 simulated datasets; all methods are applied to the same datasets at each grid value.}
\label{fig:pass48_prob_52mo}
\end{figure}

\paragraph{Sensitivity to mean-trend calibration.}
To probe whether these qualitative conclusions persist when the decision is closer to the specification limit, we repeated the decision simulation with a tighter mean trend (population mean crossing at 52 months).
Figure~\ref{fig:pass48_prob_52mo} shows that support probabilities decrease for all methods, and that the AICc step-down can be more aggressive than the known-variance reference curve in the small-variance setting.
For example, at $\mathrm{vcfrac}_{\mathrm{true}}=0.10$ the reference curve is 0.263, while SAT\_AICc yields 0.443.
This behavior is consistent with aggressive pooling when true lot-to-lot variance is small but nonzero, and it motivates caution when using automated selection for expiry decisions close to the specification limit. AICc selects the pooled model in a substantial fraction of datasets when the fitted lot variance is small but positive. Pooled OLS confidence limits then do not account for between-lot variability. The excess support probability under SAT\_AICc in Figure~\ref{fig:pass48_prob_52mo} is consistent with the undercoverage observed for SAT\_AICc in Figure~\ref{fig:coverage_tp48} in the same small-variance setting.
Appendix~\ref{app:covtest_sensitivity} adds SAT\_covtest25 to this tighter calibration. At $\mathrm{vcfrac}_{\mathrm{true}}=0.10$, SAT\_covtest25 gives Support48 of 0.326, compared with 0.263 for the known-variance reference curve, 0.277 for SAT\_reduced, and 0.443 for SAT\_AICc. Thus the p-value rule is less aggressive than AICc but can still exceed the reference curve in the same small-variance region where the coverage diagnostic is below nominal.

\section{Discussion}

\subsection{Boundary-proximal variance components and conditional-mean inference}

From a Chemistry, Manufacturing, and Controls (CMC) perspective, lot-to-lot differences in degradation slopes are often modest in well-controlled manufacturing processes, though random slopes are routinely included as a candidate structure because slope heterogeneity cannot be ruled out \emph{a priori}.
The key finding of this paper is that any fitted variance component that is small relative to residual variation---whether a random slope or a random intercept---can induce unstable contrast-specific DDF behavior under Satterthwaite---and similarly under Kenward--Roger in our sensitivity analyses---for conditional-mean (EBLUP-based) predictions.

The worked example (Figures~\ref{fig:jmp-default}--\ref{fig:sas-overlay}), the analytical concavity calculation (Section~\ref{sec:vcond_mechanism}), and the simulation mechanism plots (Figures~\ref{fig:sim_width_est}--\ref{fig:sim_df_est}) show this pattern. This behavior is operationally important because stability expiry/extension conclusions are often based on the first scheduled month at which a worst-case lot's lower confidence limit falls below specification.
When SAT produces highly variable DDF across time points (and across lots) in the small-variance setting, grid-based first-crossing times can shift earlier even when the observed measurements themselves are not close to specification limits.
Thus, the practical issue is not the exact-zero fit alone, but the unstable right-neighborhood behavior when fitted variance components are small and positive.

By contrast, the containment method yields stable DDF in the balanced stability designs studied here and avoids sharp discontinuities as fitted variance components approach the boundary.
In addition to stabilizing interval width, containment-based inference with the full random-effects model provides a single modeling framework that avoids the discontinuities introduced by data-dependent model reduction or selection rules (e.g., Q1E $p=0.25$ pooling tests, variance-contribution thresholds, or automated information-criterion selection).
We do not claim that containment is the theoretically optimal DDF for EBLUP-based conditional-mean contrasts. Rather, it is the available design-based option that is not driven by small changes in fitted variance components at a non-regular boundary. In the main simulation grid, its pointwise coverage stayed close to nominal (0.946--0.957). The modest undercoverage in the active random-slope sensitivity (minimum 0.911) indicates that no single DDF option dominates all parts of the parameter space, but containment gives the most stable operational behavior among the evaluated full-model procedures.

The instability documented here is specific to conditional-mean inference.
For marginal-mean (fixed-effect-only) predictions from the same fitted model, Satterthwaite DDF are substantially more stable near the boundary. At the worked-example profiler point shown in Figures~\ref{fig:jmp-default}--\ref{fig:sas-overlay} (Lot~G at Month~24), \texttt{PROC MIXED} reports DDF${}=13.2$ for the marginal prediction but DDF${}=1$ for the corresponding conditional prediction (Appendix~\ref{app:satt_rebuild}). However, marginal means are not the primary target for worst-case-lot expiry decisions, which require lot-specific conditional-mean bounds. Consequently, practitioners who inspect only fixed-effect or marginal-mean DDF may miss the conditional-mean SAT collapse that drives the expiry-support issue studied here.

The matched Model~(1) sensitivity results should also be interpreted in light of the prediction target. Robinson (1991) emphasized that BLUPs are predictors of realized random effects that use distributional information, not fixed-effect estimates of lot-specific parameters.\citep{robinson1991blup} The EBLUP-based conditional limits studied here therefore combine shrinkage, plug-in variance components, prediction standard errors, and approximate DDF. Exact finite-sample nominal coverage for realized lot-specific conditional means is not guaranteed simply because the fitted random-effects structure matches the data-generating structure. In the active random-slope setting, this distinction is amplified at late stability times because the random-slope contribution scales with $t_*^2$. We therefore evaluate the methods by a collection of operating characteristics: pointwise coverage, proposed-expiry support probability, stability near the boundary, and sensitivity to data-dependent model reduction.

\subsection{Model reduction, the boundary issue, and automated selection}
\label{sec:boundary_modelsel}

Testing a variance component at the null value 0 is a non-regular problem because the null lies on the boundary of the parameter space; standard likelihood theory does not apply.\citep{selfliang,Scheipl2008}
This boundary issue affects both formal tests and pragmatic procedures that behave like implicit tests of whether a variance component equals zero, including information-criterion step-down procedures.

We evaluated a simple 10\% variance-contribution reduction procedure as a pre-specified operational heuristic for SAT-only environments (e.g., base JMP).
For Model~(2), the variance-fraction interpretation is straightforward; for Model~(1), the slope-variance contribution is evaluated at the proposed expiry month $t_*$.
In Figure~\ref{fig:pass48_prob} at $\mathrm{vcfrac}_{\mathrm{true}}=0$, SAT supports the proposed expiry in 0.616 of datasets, while SAT\_reduced increases support to 0.684 by stepping down to pooled OLS in datasets where the fitted lot variance is negligible.
This mitigation comes at the cost of a threshold-based discontinuity, but it avoids the most extreme SAT behavior without relying on fully automated selection.

Information-criterion procedures such as AICc provide a complementary pragmatic alternative,\citep{Sugiura1978,HurvichTsai1989,Pack2018Stability} but they inherit the same boundary non-regularity as likelihood-based variance-component testing.\citep{selfliang,Scheipl2008}
In the baseline calibration (Figure~\ref{fig:pass48_prob}), SAT\_AICc roughly tracks the known-variance reference across much of the grid.
However, in the tighter calibration (Figure~\ref{fig:pass48_prob_52mo}) SAT\_AICc can be more aggressive than the known-variance reference curve in the small-variance setting; for example, at $\mathrm{vcfrac}_{\mathrm{true}}=0.10$ the reference-curve support probability is 0.263 while SAT\_AICc yields 0.443.
This excess support probability is consistent with the undercoverage observed for SAT\_AICc in Figure~\ref{fig:coverage_tp48}: in both cases, aggressive pooling produces confidence limits that are too narrow when true lot-to-lot variance is small but nonzero.
For this reason, we recommend treating AICc step-down as a sensitivity analysis rather than as an unqualified automated default for expiry support conclusions.

The appendix sensitivity analyses clarify the role of both the pre-specified variance-contribution rule and the p-value-based comparator. For the variance-contribution rule, 5\% and 10\% give similar conclusions, whereas 20\% is more aggressive and gives lower coverage; this supports keeping 10\% as a conservative operational compromise. A p-value-based random-effect reduction procedure using \texttt{COVTEST} and a $p=0.25$ cutoff is a natural mixed-model analog of the Q1E fixed-effect pooling heuristic, but it inherits the same boundary non-regularity as formal variance-component tests. SAT\_covtest25 is also more aggressive than SAT\_reduced in low-variance settings and can exceed the known-variance reference curve under the tighter mean-trend calibration. We therefore treat SAT\_covtest25, like AICc step-down, as an informative sensitivity analysis rather than a primary SAT-only decision rule.

The $\mathrm{vcfrac}_{\mathrm{true}}=0$ grid point in the main simulations is the Model~(3)-true case. At this point, pooled OLS tracks the known-variance reference closely, whereas unreduced mixed-model procedures can remain conservative because they retain random-effect structure at or near the variance-component boundary.
\subsection{Practical recommendation hierarchy for routine stability analyses}

For routine automated stability analytics targeting conditional-mean lower confidence bounds for observed lots, we recommend the following practical hierarchy:

\begin{enumerate}[leftmargin=*]
\item \textbf{Fit the full mixed model.} Fit the random-intercept-and-slope model (Model~(1)) using bounded REML so that conditional-mean limits are defined. In JMP, this requires changing the default variance-component setting from unbounded to bounded; see Section~\ref{sec:software}.

\item \textbf{If containment is available, use it from the full model.} In SAS \texttt{PROC MIXED} and JMP Pro (19+), use \texttt{DDFM=CONTAIN} or the analogous JMP setting to compute conditional-mean confidence limits directly from the full model. This keeps a single modeling framework and avoids discontinuities from threshold-based model switching.

\item \textbf{If containment is unavailable and Satterthwaite is the only option, use a pre-specified reduction rule.} In Satterthwaite-only environments, apply the 10\% variance-contribution reduction procedure in Section~\ref{sec:reduced}. Refit a reduced model only when the fitted random-effect contribution at the proposed expiry is negligible, and compute conditional-mean limits using SAT on the final model. The cutoff should be treated as a pre-specified operational rule, not as a sharp scientific boundary.

\item \textbf{Use AICc and p-value step-down only as sensitivity analyses.} AICc and SAT\_covtest25 step-down can be informative, but in near-boundary settings automated pooling or variance-component testing can be too aggressive for expiry-support decisions.
\end{enumerate}

When active random-slope heterogeneity is expected, or when fitted variance contributions are near a decision boundary, the operating-characteristic and threshold-sensitivity results in Appendices~\ref{app:threshold_sensitivity}--\ref{app:m1_sensitivity} should be consulted. The KR sensitivity analyses showed behavior similar to the Satterthwaite option.

\subsection{Scope and extensions}

This paper focuses on conditional-mean confidence bounds for observed lots, which align with worst-case-lot style stability decisions.
Other objectives require different methods.
Predicting future lots or controlling future lot-to-lot variability can motivate tolerance intervals or prediction intervals, which are discussed in regulatory guidance.\citep{ICHQ1Draft2025}
Out-of-specification investigations and release decisions are distinct from stability shelf-life estimation and are outside the scope of this work.

The closed-form analytical results in Section~\ref{sec:vcond_mechanism} and Appendix~\ref{app:closedform} are intentionally derived for the balanced random-intercept design. The specific expressions for $\vcond$, $\nu_\Delta$, and the right-hand boundary limit do not extend unchanged to unbalanced pull schedules or richer random-effect covariance structures, where the analytical signature is design-dependent. The qualitative mechanism, however---first-order propagation through a bounded, concave prediction-variance function near a variance-component boundary---is not specific to that special case. The worked example, the primary simulations, the full Model~(1) fits, the matched Model~(1) sensitivity, and the KR sensitivity analyses all show the same near-boundary collapse pattern. For substantially unbalanced designs, analysts should treat the closed form as a diagnostic explanation and rely on design-specific operating-characteristic checks for the final procedural assessment.

\section{Conclusion}

When a fitted random-effect variance component is close to the boundary at zero, variance-component-dependent DDF methods can become unstable for conditional-mean predictions in random-lot stability mixed models. In the settings studied here, Satterthwaite DDF---and Kenward--Roger DDF in sensitivity analyses---can collapse, inflating $t$ critical values and yielding very wide and sometimes nonmonotone pointwise confidence limits on scheduled time grids. In the balanced designs studied here, this behavior can change proposed-expiry support conclusions even when the observed measurements themselves are not close to specification limits.

Containment-based inference with the full random-effects model is the preferred routine workflow when available, because it avoids boundary-driven DDF discontinuities without requiring threshold-based model reduction. In SAT-only environments, the pre-specified 10\% variance-contribution reduction rule provides a pragmatic mitigation by simplifying only when fitted random-effect contributions at the proposed expiry are negligible. AICc and COVTEST step-down procedures are best treated as sensitivity analyses rather than primary workflows, especially when the proposed-expiry margin is small.

As mixed-model methods become more prominent in regulatory stability frameworks, explicit guidance on the DDF method for conditional-mean inference would help ensure consistency across software platforms and across submissions.

\section*{Statements}

\noindent\textbf{Data availability.} The worked example dataset and simulation files supporting the findings are available in the data repository.\citep{KarlMendeleyData2025}

\noindent\textbf{Code availability.} Simulation scripts and run settings are available in the data repository.\citep{KarlMendeleyData2025}

\noindent\textbf{Conflict of interest.} The authors declare no conflicts of interest.

\noindent\textbf{Funding.} No external funding.

\noindent\textbf{Ethics approval.} Not applicable.

\noindent\textbf{Consent to participate.} Not applicable.

\noindent\textbf{Consent for publication.} Not applicable.

\section*{Acknowledgments}
The first author used GPT-5.5 Pro for proofreading/rewording, LaTeX formatting, and assistance in structuring and commenting the SAS and R simulation code based on the author's instructions. The tool was used to improve clarity and streamline formatting and code documentation. In the closed-form derivation (Appendix~\ref{app:closedform}), GPT-5.5 Pro suggested specific technical steps including the orthogonal rotation of lot effects and the Schur-complement reduction of the mixed-model coefficient matrix; the first author re-derived each step and verified the resulting closed-form covariance and DDF expressions numerically against the \texttt{PROC MIXED} \texttt{AsyCov} table and the finite-difference reconstruction in Appendix~\ref{app:satt_rebuild}. All authors reviewed and approved the manuscript and accept full responsibility for its content.

\appendix

\section{Baseline calculations under known variance components}
\label{app:baseline}
This appendix provides reference calculations used in the main paper when the data-generating random-intercept Model~(2) is treated as known (i.e., variance components are fixed at their generating values rather than estimated).
Appendix~\ref{app:tref_lsl} derives a pooled-regression reference crossing time for the special case $\sigma_{b0}^2=0$.
Appendix~\ref{app:vc_known_ci} defines the conditional-mean prediction-error variance $V_{\mathrm{CI}}(t_*)$, a shared building block for the remaining results.
Using $V_{\mathrm{CI}}(t_*)$, Appendix~\ref{app:baselinepass_lsl} defines the reference probability that all lots pass the decision criterion at the proposed expiry month $t_*=48$ (``Expected'' in Figure~\ref{fig:pass48_prob}).
This known-variance reference is the pass probability of the conditional-mean lower confidence bound decision rule at $t_*$ for the observed lots, not the population probability that a randomly selected lot's true mean $\mu_i(t_*)$ exceeds the LSL. It omits variance-component estimation, boundary truncation, DDF approximation, and model-selection effects.
Throughout this appendix, Normal critical values are used because variance components are treated as known.

\subsection{Pooled-regression reference crossing time ($\sigma_{b0}^2=0$)}
\label{app:tref_lsl}

When $\sigma_{b0}^2=0$, Model~(2) reduces to pooled OLS (Model~(3)).
Write the true pooled mean as $m(t)=\beta_0+\beta_1 t$.
Using a Normal critical value (variance treated as known), a one-sided $100(1-\alpha)\%$ lower bound for the pooled mean at time $t$ is
\[
\mathrm{LCL}_{\mathrm{OLS}}(t)
=
 m(t)
-
 z_{1-\alpha}\ \sigma_e\
\sqrt{\frac{1}{n}+\frac{(t-\bar t)^2}{S_{xx}}},
\]
where $\bar t$ and $S_{xx}$ are the usual design constants from pooled regression.
Define the pooled-regression reference crossing time $t_{\mathrm{ref}}$ by $\mathrm{LCL}_{\mathrm{OLS}}(t_{\mathrm{ref}})=\mathrm{LSL}$.

Under the calibration used in the simulation study, $\beta_1=(\mathrm{LSL}-\beta_0)/57=-10/57$, so $m(48)=91.58>\mathrm{LSL}$ and solving $\mathrm{LCL}_{\mathrm{OLS}}(t)=\mathrm{LSL}$ with $\alpha=0.05$ yields $t_{\mathrm{ref}}\approx 53.04$ months (beyond the 48-month decision horizon).
For the balanced simulation design, $L=10$ lots are observed at months $t\in\{0,3,6,9,12,24,36\}$ (7 observations per lot), for a total of $n=70$ observations; for this design, $\bar t=90/7\approx 12.86$ and $S_{xx}=68940/7\approx 9848.57$.

\subsection{Conditional-mean prediction-error variance under the random-intercept model}
\label{app:vc_known_ci}

For $\mathrm{vcfrac}_{\mathrm{true}}\in(0,1)$, write the data-generating process as the random-intercept Model~(2)
\[
Y_{ij}=\beta_0+\beta_1 t_j + b_i + \varepsilon_{ij},\qquad
b_i\sim N(0,\sigma_{b0}^2),\quad \varepsilon_{ij}\sim N(0,\sigma_e^2),
\]
with $\sigma_{b0}^2=\mathrm{vcfrac}_{\mathrm{true}}\, \sigma_{\mathrm{tot}}^2$ and $\sigma_e^2=(1-\mathrm{vcfrac}_{\mathrm{true}})\, \sigma_{\mathrm{tot}}^2$.
Fix an evaluation time $t_*$ and define the lot-specific conditional mean
\[
\mu_i(t_*)=\beta_0+\beta_1 t_*+b_i.
\]

Let $y$ be the stacked response vector with fixed-effects design matrix $X$ (intercept and time) and random-effects design matrix $Z$ (lot indicators for the random intercept).
With $G=\sigma_{b0}^2 I_L$ and $R=\sigma_e^2 I_n$, the marginal covariance is $V=ZGZ^\top+R$.
Treating $(\sigma_{b0}^2,\sigma_e^2)$ as known at their generating values, the BLUP/GLS predictor $\widehat{\mu}_i(t_*)$ is a linear function of $y$ and its prediction error is Normal with variance
\[
V_{\mathrm{CI}}(t_*)=\mathrm{Var}\{\widehat{\mu}_i(t_*)-\mu_i(t_*)\}.
\]

We compute $V_{\mathrm{CI}}(t_*)$ exactly using Henderson's mixed model equations.\citep{henderson1959}
Let $x_*=(1,t_*)^\top$ denote the fixed-effects design row at $t_*$ and let $e_i$ denote the $i$th standard basis vector for lot $i$ in the random-intercept design.
If $C$ is the inverse of the mixed-model coefficient matrix for $(\widehat\beta,\widehat b)$, then
\[
V_{\mathrm{CI}}(t_*)=
\begin{pmatrix}
x_*\\ e_i
\end{pmatrix}^{\!\top}
C
\begin{pmatrix}
x_*\\ e_i
\end{pmatrix}.
\]
Because the design is balanced and exchangeable across lots under the random-intercept model, $V_{\mathrm{CI}}(t_*)$ does not depend on $i$.

\subsection{Known-variance all-lots pass probability under the random-intercept model}
\label{app:baselinepass_lsl}

Fix the proposed expiry month $t_*=48$.
Using a one-sided Normal critical value and the prediction-error variance $V_{\mathrm{CI}}(48)$ from Appendix~\ref{app:vc_known_ci}, define the reference lower bound at $t_*$ as
\[
\mathrm{LCL}_i(t_*)=\widehat\mu_i(t_*)-z_{1-\alpha}\sqrt{V_{\mathrm{CI}}(t_*)}.
\]
The all-lots pass event is $\mathrm{LCL}_i(t_*)\ge \mathrm{LSL}$ for every lot $i$, equivalently
\[
\widehat\mu_i(t_*) \ge \mathrm{LSL}+z_{1-\alpha}\sqrt{V_{\mathrm{CI}}(t_*)}
\quad\text{for all } i=1,\ldots,L,
\]
with $L=10$.
Note that $\widehat\mu_i(t_*)$ is a predictor based on data observed through the last pull time, so the pass probability depends on both $\sigma_{b0}^2$ and $\sigma_e^2$ through $V_{\mathrm{CI}}(t_*)$ and the joint distribution of $(\widehat\mu_1(t_*),\ldots,\widehat\mu_L(t_*))$.

To evaluate this probability, we treat the vector $(\widehat\mu_1(t_*),\ldots,\widehat\mu_L(t_*))$ as multivariate Normal under the known-variance-component model and compute the corresponding multivariate Normal tail probability.
In the accompanying code, we compute the exact $10\times 10$ covariance matrix implied by the mixed model equations and evaluate the tail probability numerically using \texttt{mvtnorm::pmvnorm} in R.\citep{mvtnormManual}

\section{Satterthwaite DDF reconstruction for the worked example}
\label{app:satt_rebuild}

We reconstruct the Satterthwaite DDF for the worked example dataset (14~lots,
random-intercept Model~(2)), focusing on the profiler point used in
Figures~\ref{fig:jmp-default}--\ref{fig:sas-overlay} (Lot~G, Month~24). The graphical worked-example displays in Figures~\ref{fig:jmp-default}--\ref{fig:sas-overlay} use two-sided 95\% intervals for visual illustration. The numerical interval values below follow the formal one-sided 95\% lower-limit convention, i.e., two-sided 90\% \texttt{PROC MIXED} prediction limits with \texttt{ALPHAP=0.10}.
Using \texttt{PROC MIXED}, we request both conditional (\texttt{OUTP=}) and
marginal (\texttt{OUTPM=}) predictions at this scored point.
We then rebuild Satterthwaite DDF via Eq.~\eqref{eq:satt_df_delta} by extracting
$\widehat{\bm\Omega}$ from the \texttt{AsyCov} table and computing
$g(\widehat\theta)=\partial \widehat v/\partial\theta$ by central finite differences
with covariance parameters held fixed via \texttt{PARMS}/\texttt{HOLD} (no-iteration
refits).
Here $\widehat v$ denotes the prediction-error variance reported by \texttt{PROC MIXED}
(\texttt{StdErrPred}$^2$).

The REML covariance parameter estimates are
$\hat\theta_1=\hat\sigma_{b0}^2=0.004491$ and
$\hat\theta_2=\hat\sigma_e^2=0.2360$, giving
$\widehat{\mathrm{vcfrac}}\approx 0.019$.
The estimated asymptotic covariance matrix of
$\hat{\bm\theta}=(\hat\theta_1,\hat\theta_2)$ is
\[
\hat{\bm\Omega}=
\begin{pmatrix}
1.57\times 10^{-4} & -1.11\times 10^{-4}\\
-1.11\times 10^{-4} & 1.00\times 10^{-3}
\end{pmatrix}.
\]

Table~\ref{tab:satt_rebuild} compares conditional and marginal predictions from the
same fit.
The two point predictions differ because the conditional prediction includes the EBLUP
for Lot~G (a fitted random intercept), whereas the marginal prediction equals
$x(t)^\top\hat{\bm\beta}$.
The marginal gradient component
$\hat g_1=\partial \hat v/\partial\sigma_{b0}^2$ is small (0.071 in this example),
while the conditional gradient is nearly ten times larger (0.677).
Within the delta-method formula~\eqref{eq:satt_df_delta}, this amplified gradient is
the driver of the DDF collapse: it inflates
$\hat{\bm g}^\top\hat{\bm\Omega}\hat{\bm g}$ by a factor of about 96 relative to
the marginal case, producing a very small denominator degrees of freedom.
The rebuilt DDF match the SAS-reported values to numerical precision for
\texttt{OUTPM}.
For \texttt{OUTP}, the delta-method reconstruction yields $\nu=0.84$, while
\texttt{PROC MIXED} reports the operational DDF as $\nu=1.00$ and uses the
corresponding $t_{0.95,1}=6.31$ multiplier. This small discrepancy occurs in
the extreme near-boundary range and does not affect the diagnosis: both
calculations show that the conditional DDF has collapsed to roughly one degree
of freedom, with the software-reported value slightly less extreme than the raw
delta-method reconstruction.

The closed-form calculation in Section~\ref{sec:vcond_mechanism} shows that the conditional prediction-error variance $v_{\mathrm{cond}}(\sigma_{b0}^2,\sigma_e^2)$ is a concave, saturating function of the lot variance component in the balanced random-intercept design.
Thus the delta-method approximation in Eq.~\eqref{eq:satt_df_delta} is not merely using a large gradient; it is propagating uncertainty through a first-order tangent approximation to a nonlinear function at a point very close to the boundary.

\begin{table}[htbp]
\centering
\caption{Satterthwaite DDF reconstruction for the worked example (Lot~G, Month~24,
Model~(2)).}
\label{tab:satt_rebuild}
\small
\begin{tabular}{lcc}
\toprule
Quantity & \texttt{OUTP} (conditional) & \texttt{OUTPM} (marginal) \\
\midrule
Prediction $\hat\mu$ & 100.037 & 100.025 \\
$\hat v = \mathrm{SE}^2$ & 0.005451 & 0.002212 \\
$\hat g_1\;(\partial\hat v/\partial\sigma_{b0}^2)$
  & 0.677 & 0.071 \\
$\hat g_2\;(\partial\hat v/\partial\sigma_e^2)$
  & 0.0102 & 0.0080 \\
$\hat{\bm g}^\top\hat{\bm\Omega}\,\hat{\bm g}$
  & $7.07\times 10^{-5}$ & $7.40\times 10^{-7}$ \\
SAS-reported DDF $\nu$ & 1.00 & 13.22 \\
Rebuilt DDF $\nu$ & 0.84 & 13.22 \\
Implied $t_{0.95,\nu}$ multiplier (\texttt{PROC MIXED}) & 6.31 & 1.77 \\
\bottomrule
\end{tabular}
\end{table}

Table~\ref{tab:kr_worked_example} gives the corresponding \texttt{PROC MIXED} Kenward--Roger worked-example output at the same scoring point. We report this as a software evaluation rather than a full KR reconstruction, because the KR-adjusted covariance and prediction-variance components are not exposed by \texttt{PROC MIXED} in the same way as the Satterthwaite delta-method quantities. At the conditional scoring point, KR reports DDF=1 and the same one-sided multiplier as SAT. Because the KR prediction standard error is larger, the resulting interval is wider than the SAT interval.

\begin{table}[htbp]
\centering
\caption{Kenward--Roger worked-example evaluation for the Lot~G, Month~24 scoring point.}
\label{tab:kr_worked_example}
\small
\begin{threeparttable}
\begin{tabular}{llrrrrr}
\toprule
Method & Prediction & DDF & SE & LCL & UCL & $t_{0.95,\nu}$ \\
\midrule
CONTAIN & Conditional & 111.00 & 0.0738 & 99.915 & 100.160 & 1.659 \\
SAT & Conditional & 1.00 & 0.0738 & 99.571 & 100.503 & 6.314 \\
KR & Conditional & 1.00 & 0.1122 & 99.329 & 100.746 & 6.314 \\
CONTAIN & Marginal & 111.00 & 0.0470 & 99.947 & 100.103 & 1.659 \\
SAT & Marginal & 13.22 & 0.0470 & 99.942 & 100.108 & 1.769 \\
KR & Marginal & 13.22 & 0.0470 & 99.942 & 100.108 & 1.769 \\
\bottomrule
\end{tabular}
\begin{tablenotes}
\footnotesize
\item Note. Values are for Lot~G at Month~24 in the 14-lot, 9-pull worked example. LCL/UCL are the two-sided 90\% \texttt{OUTP}/\texttt{OUTPM} limits corresponding to one-sided 95\% lower confidence limits. KR denotes \texttt{DDFM=KR2} output from \texttt{PROC MIXED}.
\end{tablenotes}
\end{threeparttable}
\end{table}

SAS code and outputs for the reconstruction are available in the data repository.\citep{KarlMendeleyData2025}

\section{Closed-form conditional prediction variance and delta-method DDF in the balanced random-intercept design}
\label{app:closedform}
\label{app:vcond_derivation}

This appendix derives the closed form of $V_{\mathrm{CI}}(t_*)$ from Appendix~\ref{app:vc_known_ci} in the balanced random-intercept design and extends it to the delta-method Satterthwaite DDF. For compactness we write $\tau=\sigma_{b0}^2$, $s=\sigma_e^2$, and $x_*=(1,t_*)^\top$, and we denote the resulting variance function by $\vcond(t_*;\tau,s)$ when emphasizing its dependence on the variance components.
The decomposition below is a specialization of the standard mixed-model-equation and BLUP prediction-variance framework to the balanced random-intercept design at a specified covariate point \citep{henderson1959,robinson1991blup,SearleCasellaMcCulloch1992}.
We give the derivation because the explicit closed form is the input to the delta-method denominator degrees of freedom (DDF) calculation and to the boundary limit.
The derivation has three pieces. First, the conditional-mean prediction-error variance is written as a quadratic form from the mixed-model equations and evaluated in closed form by rotating the $L$ lot effects into the average, or grand-lot, random-effect direction and the remaining $L-1$ orthogonal lot-contrast directions. Second, the balanced REML covariance matrix for $(\widehat\tau,\widehat s)$ is written using standard balanced one-way variance-component results. Third, these pieces are combined through the delta-method Satterthwaite formula to obtain the closed-form DDF, its variance-fraction form, and its right-hand boundary limit as $\tau\downarrow0$.

\paragraph{Prediction-error variance from the mixed-model equations.}
Consider the balanced random-intercept model
\[
Y_{ij}=x_j^\top\beta+b_i+\varepsilon_{ij},\qquad
b_i\sim N(0,\tau),\quad \varepsilon_{ij}\sim N(0,s),
\]
with $L$ lots, $m$ common time points per lot, and $x_j=(1,t_j)^\top$.
Let $\widetilde X$ be the $m\times 2$ within-lot fixed-effect design matrix. The full fixed-effect design is $X=\mathbf 1_L\otimes \widetilde X$, so $X^\top X=L\widetilde X^\top\widetilde X$. Let $Z=I_L\otimes\mathbf 1_m$ denote the $Lm\times L$ lot-indicator matrix.
For the target $\mu_i(t_*)=x_*^\top\beta+b_i$, the mixed-model-equation prediction-error variance is
\[
\operatorname{Var}\{\widehat\mu_i(t_*)-\mu_i(t_*)\}=c_i^\top C c_i,
\qquad
c_i=\begin{pmatrix}x_*\\ e_i\end{pmatrix},
\]
where $e_i$ is the $i$th standard basis vector in $\mathbb R^L$, and $C$ is the inverse of the mixed-model coefficient matrix
\[
\begin{pmatrix}
X^\top R^{-1}X & X^\top R^{-1}Z\\
Z^\top R^{-1}X & Z^\top R^{-1}Z+G^{-1}
\end{pmatrix},
\qquad
R=sI_{Lm},\quad G=\tau I_L.
\]
$C$ is the joint prediction-error covariance matrix for $(\widehat\beta,\widehat b-b)$ \citep{henderson1959,SearleCasellaMcCulloch1992}.

To evaluate the quadratic form, rotate the lot effects into the grand-lot direction, proportional to $\mathbf 1_L$, and $L-1$ orthogonal lot contrasts. Let $T=(L^{-1/2}\mathbf 1_L,T_c)$ be an orthogonal $L\times L$ matrix with the columns of $T_c$ spanning the subspace orthogonal to $\mathbf 1_L$, and write $\widetilde b=T^\top b$. The grand-lot coordinate is therefore the normalized average-direction coordinate
\[
\widetilde b_1=L^{-1/2}\mathbf 1_L^\top b,
\]
not the arithmetic average $L^{-1}\mathbf 1_L^\top b$. Because $G=\tau I_L$, the rotated effects satisfy
\[
\operatorname{Var}(T^\top b)=T^\top(\tau I_L)T=\tau I_L,
\]
so the orthogonal rotation leaves the model unchanged and merely separates the grand-lot direction from the $L-1$ lot-contrast directions.

The rotation also preserves the quadratic form: $c_i^\top C c_i=\widetilde c_i^{\,\top}\widetilde C\,\widetilde c_i$, where $\widetilde C$ is the inverse of the correspondingly rotated coefficient matrix and $\widetilde c_i=\bigl(x_*^\top,(T^\top e_i)^\top\bigr)^\top$ has grand-lot component $L^{-1/2}$ and contrast components $T_c^\top e_i$.
Since each lot has the same within-lot design,
\[
X^\top Z=(\mathbf 1_L^\top\otimes \widetilde X^\top)(I_L\otimes\mathbf 1_m)
=\mathbf 1_L^\top\otimes u,
\qquad u=\widetilde X^\top\mathbf 1_m,
\]
so $X^\top Z$ is a $2\times L$ matrix with every column equal to $u$. Multiplication by the first column of $T$ sums these equal columns and multiplication by any contrast column of $T_c$ gives zero, yielding
\[
X^\top ZT=\bigl(\sqrt L\,u,\;0,\ldots,0\bigr).
\]
Thus the $L-1$ lot-contrast coordinates decouple from the fixed effects. Equivalently, after rotation the relevant mixed-model coefficient matrix separates into one block for the fixed effects and the grand-lot coordinate, and another for the $L-1$ lot contrasts.
Also,
\[
T^\top(Z^\top R^{-1}Z+G^{-1})T
=\left(\frac{m}{s}+\frac1\tau\right)I_L,
\]
because $Z^\top Z=mI_L$. Hence each contrast coordinate has mixed-model-equation precision $m/s+1/\tau$ and prediction-error variance $1/(m/s+1/\tau)=s\tau/(s+m\tau)$.
The contrast components of $\widetilde c_i$ have squared length $\|T_c^\top e_i\|^2=e_i^\top T_cT_c^\top e_i=1-1/L=(L-1)/L$; equivalently, $e_i-L^{-1}\mathbf 1_L$ is the projection of $e_i$ onto the $(L-1)$-dimensional contrast subspace. Therefore the total lot-contrast contribution is
\[
\frac{L-1}{L}\frac{s\tau}{s+m\tau}.
\]

It remains to account for the fixed effects and the grand-lot direction. The coefficient block for $(\beta,\widetilde b_1)$ is
\[
\widetilde M_{\mathrm{top}}=
\begin{pmatrix}
s^{-1}X^\top X & s^{-1}\sqrt L\,u\\
s^{-1}\sqrt L\,u^\top & m/s+1/\tau
\end{pmatrix},
\]
with scoring vector $\widetilde c=(x_*^\top,L^{-1/2})^\top$, the top block of $\widetilde c_i$.
Let $A=s^{-1}X^\top X$, $B=s^{-1}\sqrt L\,u$, $D=m/s+1/\tau$, and $S=D-B^\top A^{-1}B$.
Using $(s^{-1}X^\top X)^{-1}=s(X^\top X)^{-1}$ and $(X^\top X)^{-1}=L^{-1}(\widetilde X^\top\widetilde X)^{-1}$, the remaining identity
\[u^\top(\widetilde X^\top\widetilde X)^{-1}u=m
\]
follows because $u=\widetilde X^\top\mathbf 1_m$ and $\mathbf 1_m$ lies in the column space of $\widetilde X$ through the intercept column. Equivalently,
\[u^\top(\widetilde X^\top\widetilde X)^{-1}u
=
\mathbf 1_m^\top
\widetilde X(\widetilde X^\top\widetilde X)^{-1}
\widetilde X^\top\mathbf 1_m
=
\mathbf 1_m^\top\mathbf 1_m
=
m .
\]
Thus the Schur complement is
\[
S=m/s+1/\tau
- s^{-2}L\,u^\top\{s(X^\top X)^{-1}\}u
= m/s+1/\tau - m/s
=1/\tau.
\]
For a block matrix $\begin{pmatrix}A&B\\B^\top&D\end{pmatrix}$ with scalar lower-right block $D$ and Schur complement $S=D-B^\top A^{-1}B$, the inverse quadratic form satisfies
\[
\begin{pmatrix}a\\d\end{pmatrix}^\top
\begin{pmatrix}A&B\\B^\top&D\end{pmatrix}^{-1}
\begin{pmatrix}a\\d\end{pmatrix}
= a^\top A^{-1}a
+ S^{-1}\{d-B^\top A^{-1}a\}^2.
\]
Applying this identity with $a=x_*$ and $d=L^{-1/2}$ gives
\[
\widetilde c^\top\widetilde M_{\mathrm{top}}^{-1}\widetilde c
= s x_*^\top(X^\top X)^{-1}x_* + \frac{\tau}{L}
  \left\{x_*^\top(\widetilde X^\top\widetilde X)^{-1}u-1\right\}^2.
\]
Writing $q=x_*^\top(\widetilde X^\top\widetilde X)^{-1}u$, this is $s x_*^\top(X^\top X)^{-1}x_*+\tau(q-1)^2/L$.
Because the model includes an intercept and the time points are common across lots,
\[
(\widetilde X^\top\widetilde X)^{-1}u
=
(\widetilde X^\top\widetilde X)^{-1}\widetilde X^\top\mathbf 1_m
=
e_1,
\]
where \(e_1=(1,0)^\top\) is the intercept basis vector in \(\mathbb R^2\). Therefore, with \(x_*=(1,t_*)^\top\),
\[
q=x_*^\top e_1=1.
\]
It follows that the additional grand-lot contribution,
\[
\frac{\tau}{L}(q-1)^2,
\]
is identically zero. This cancellation occurs because the fixed-effect space contains the intercept direction and each lot has the same time design. The prediction time \(t_*\) still affects the fixed-effect leverage term
\[
x_*^\top(X^\top X)^{-1}x_*,
\]
but it does not affect \(q\), which only records the intercept coordinate of \(x_*\).

Combining this cancellation with the lot-contrast component yields
\[
\operatorname{Var}\{\widehat\mu_i(t_*)-\mu_i(t_*)\}
=
s\,x_*^\top(X^\top X)^{-1}x_*
+
\frac{L-1}{L}\frac{s\tau}{s+m\tau},
\]
which is Eq.~\eqref{eq:vcond_closed}. Thus, in balanced data, the conditional prediction variance has two pieces: a fixed-effect leverage term, obtained after cancellation of the grand-lot contribution, and a shrinkage-corrected lot-contrast term that is increasing, concave, and saturating in \(\tau\).

Let
\[
A_* = x_*^\top(X^\top X)^{-1}x_*.
\]
Differentiating the closed form gives
\[
g_\tau=\frac{\partial \vcond}{\partial\tau}
=\frac{s^2(L-1)}{L(s+m\tau)^2},
\qquad
 g_s=\frac{\partial \vcond}{\partial s}
=A_*+\frac{m(L-1)\tau^2}{L(s+m\tau)^2}.
\]
The gradient has a nonzero positive-side boundary limit:
$g_\tau\to (L-1)/L$ and $g_s\to A_*$ as $\tau\downarrow0$.
The derivative $g_\tau$ is largest at the boundary and decreases as $(s+m\tau)^{-2}$. Equivalently, anticipating the variance-fraction parametrization introduced below, if $p=\tau/(\tau+s)$ and $s$ is held fixed, the lot-contrast term is
\[
\frac{s(L-1)}{L}\frac{p}{1+(m-1)p},
\]
whose derivative with respect to $p$ is proportional to $\{1+(m-1)p\}^{-2}$.

\paragraph{Balanced REML covariance matrix.}
For the same balanced random-intercept design, the interior REML covariance matrix, corresponding to the covariance-parameter \texttt{AsyCov} matrix used in the delta-method calculation, can be written in closed form.
Let $r_W$ denote the dimension of the within-lot fixed-effect design after removing the intercept; equivalently,
\[
r_W=\operatorname{rank}\left\{\left(I_m-m^{-1}\mathbf 1_m\mathbf 1_m^\top\right)\widetilde X\right\},
\qquad
 d_W=L(m-1)-r_W.
\]
In this balanced random-intercept model with $n=Lm$ observations,
\[
n-\operatorname{rank}([X\ Z])
=
Lm-(L+r_W)
=
L(m-1)-r_W
=
d_W,
\]
which is the containment DDF under the retained random-intercept structure
in this balanced setting.
For the standard stability model with $\widetilde X=(\mathbf 1_m,t)$ and common time points, only the time direction remains after removing the intercept, so $r_W=1$.
In the balanced design, the interior REML estimators can be described through two independent mean squares: a between-lot mean square $\mathrm{MS}_B$ with $L-1$ degrees of freedom and expectation $s+m\tau$, and a within-lot mean square $\mathrm{MS}_W$ with $d_W$ degrees of freedom and expectation $s$. Away from the boundary, $\widehat s=\mathrm{MS}_W$ and $\widehat\tau=(\mathrm{MS}_B-\mathrm{MS}_W)/m$. The chi-squared mean-square variances and independence give
\[
\operatorname{Var}(\mathrm{MS}_W)=\frac{2s^2}{d_W},
\qquad
\operatorname{Var}(\mathrm{MS}_B)=\frac{2(s+m\tau)^2}{L-1},
\qquad
\operatorname{Cov}(\mathrm{MS}_B,\mathrm{MS}_W)=0.
\]
Therefore, using the standard balanced one-way variance-component results \citep{SearleCasellaMcCulloch1992},
\begin{align}
\Omega_{\tau\tau}
&=\operatorname{Var}(\widehat\tau)
=\frac{1}{m^2}\{\operatorname{Var}(\mathrm{MS}_B)+\operatorname{Var}(\mathrm{MS}_W)\}
=\frac{2}{m^2}\left\{\frac{(s+m\tau)^2}{L-1}+\frac{s^2}{d_W}\right\},\nonumber\\
\Omega_{ss}
&=\operatorname{Var}(\widehat s)
=\frac{2s^2}{d_W},\nonumber\\
\Omega_{\tau s}
&=\operatorname{Cov}(\widehat\tau,\widehat s)
=\operatorname{Cov}\{(\mathrm{MS}_B-\mathrm{MS}_W)/m,\mathrm{MS}_W\}
=-\frac{2s^2}{m d_W}.
\label{eq:omega_closed}
\end{align}
At the worked-example values $L=14$, $m=9$, $r_W=1$, $\widehat s=0.2360$, and $\widehat\tau=0.004491$, Eq.~\eqref{eq:omega_closed} gives
\[
\widehat\Omega_{\tau\tau}=1.57\times 10^{-4},\qquad
\widehat\Omega_{ss}=1.00\times 10^{-3},\qquad
\widehat\Omega_{\tau s}=-1.11\times 10^{-4},
\]
matching the \texttt{PROC MIXED} \texttt{AsyCov} values in Appendix~\ref{app:satt_rebuild} to the reported precision.
At this point, both the prediction variance and the covariance of $(\widehat\tau,\widehat s)$ are in closed form, so the delta-method DDF can also be written in closed form.

\paragraph{Variance-fraction parametrization.}
Let $p=\tau/(\tau+s)$, so $\tau=sp/(1-p)$, and define
\[
c=\frac{L-1}{L},
\qquad
D(p)=1+(m-1)p,
\qquad
f(p)=A_*+c\frac{p}{D(p)}.
\]
Then Eq.~\eqref{eq:vcond_closed} becomes
\[
\vcond(t_*;s,p)=s f(p),
\]
with gradient with respect to the independent coordinates $(s,p)$
\[
g_{(s,p)}=
\begin{pmatrix}
\partial\vcond/\partial s\\[1mm]
\partial\vcond/\partial p
\end{pmatrix}
=
\begin{pmatrix}
f(p)\\[1mm]
sc/D(p)^2
\end{pmatrix}.
\]
Although the right-hand boundary limit below is most direct in $(\tau,s)$ coordinates, the variance-fraction coordinates $(s,p)$ separate residual scale from the random-intercept fraction and make transparent the opposite endpoint $p\uparrow1$, at which the lot variance dominates. We therefore transform the covariance matrix for $(\widehat\tau,\widehat s)$ to the corresponding covariance matrix for $(\widehat s,\widehat p)$, so that the same delta-method DDF can be written on the variance-fraction scale used in the figures and reduction rule.
The first-order delta-method covariance transformation uses the Jacobian
\[
J
=
\frac{\partial(s,p)}{\partial(\tau,s)}
=
\begin{pmatrix}
0 & 1\\[1mm]
(1-p)^2/s & -p(1-p)/s
\end{pmatrix}.
\]
Thus the covariance matrix for $(\widehat s,\widehat p)$, to first order, is
\[
\Omega^{(s,p)}=J\Omega^{(\tau,s)}J^\top .
\]
Substituting Eq.~\eqref{eq:omega_closed} and simplifying gives
\begin{align}
\Omega^{(s,p)}_{ss}
&=\frac{2s^2}{d_W},\nonumber\\
\Omega^{(s,p)}_{sp}
&=-\frac{2s(1-p)D(p)}{m d_W},\nonumber\\
\Omega^{(s,p)}_{pp}
&=\frac{2(1-p)^2D(p)^2}{m^2}
\left(
\frac{1}{L-1}+\frac{1}{d_W}
\right).
\label{eq:omega_sp}
\end{align}
These expressions are the same first-order (delta-method) covariance written on the variance-fraction scale used for $\mathrm{vcfrac}$.

\paragraph{Delta-method Satterthwaite DDF and boundary limit.}
Write $\theta=(\tau,s)^\top$, $v(\theta)=\vcond(t_*;\tau,s)$, and $\widehat v=v(\widehat\theta)$.
The Satterthwaite formula separates two ideas. First, it uses a scaled-$\chi^2$ moment match for the plug-in prediction variance: if $\widehat v$ were approximated as $v\,\chi^2_\nu/\nu$, then $\operatorname{Var}(\widehat v)=2v^2/\nu$; solving for $\nu$ and replacing $v$ and $\operatorname{Var}(\widehat v)$ by estimates gives
\[
\nu=\frac{2\widehat v^2}{\widehat{\operatorname{Var}}(\widehat v)}.
\]
Second, the estimate $\widehat{\operatorname{Var}}(\widehat v)$ must be supplied. The usual delta-method implementation linearizes $v(\theta)$ around $\widehat\theta$ and takes $\widehat{\operatorname{Var}}(\widehat v)=g(\widehat\theta)^\top\widehat\Omega\,g(\widehat\theta)$, where $g=\partial v/\partial\theta$. Thus the derivative calculation is one ingredient, but the DDF collapse is governed by the moment-matching ratio itself when the propagated uncertainty in $\widehat v$ is large relative to $\widehat v^2$.
Combining $g=(g_\tau,g_s)^\top$ with Eq.~\eqref{eq:omega_closed}, the balanced-design delta-method Satterthwaite DDF is
\begin{equation}
\nu_\Delta
=
\frac{2\vcond(t_*;\tau,s)^2}
{g_\tau^2\Omega_{\tau\tau}+2g_\tau g_s\Omega_{\tau s}+g_s^2\Omega_{ss}}.
\label{eq:satt_closed_balanced}
\end{equation}
Equivalently, using the $(s,p)$ gradient and Eq.~\eqref{eq:omega_sp}, the same DDF is
\begin{equation}
\nu_\Delta(p)
=
\frac{f(p)^2}
{
\displaystyle
\frac{f(p)^2}{d_W}
-
\frac{2f(p)c(1-p)}{m d_W D(p)}
+
\frac{c^2(1-p)^2}{m^2D(p)^2}
\left(
\frac{1}{L-1}+\frac{1}{d_W}
\right)
}.
\label{eq:satt_closed_p}
\end{equation}
As $p\uparrow1$, the terms multiplied by $(1-p)$ and $(1-p)^2$ vanish, leaving only $f(p)^2/d_W$ in the denominator and giving $\nu_\Delta(p)\to d_W$.
For the worked-example values, $\vcond=0.005451$ and $g^\top\Omega g=7.07\times10^{-5}$, so Eq.~\eqref{eq:satt_closed_balanced} gives
\[
\nu_\Delta=\frac{2(0.005451)^2}{7.07\times10^{-5}}=0.84,
\]
the same reconstructed value reported in Appendix~\ref{app:satt_rebuild}.
This first-order calculation identifies the Satterthwaite mechanism used in the software reconstruction. The Satterthwaite calculation supplies $\widehat{\operatorname{Var}}(\widehat v)$ by an unrestricted first-order linearization of a nonlinear, saturating covariance-parameter map evaluated adjacent to the boundary. When that propagated uncertainty is large relative to $\widehat v^2$, the scaled-$\chi^2$ moment-matching ratio translates it directly into a small DDF. The propagation diagnostic in Figure~\ref{fig:vcond_concavity}, Panel B, uses the same ingredients as Eq.~\eqref{eq:satt_closed_balanced}: the worked-example $\widehat\theta$, the covariance-parameter \texttt{AsyCov} matrix $\widehat\Omega$, the gradient $g(\widehat\theta)$, and the closed-form map $\vcond(\theta)$.

Equation~\eqref{eq:satt_closed_balanced} also shows why the collapse is design-dependent.
At the right-hand boundary limit, $\vcond\to sA_*$, $g_\tau\to (L-1)/L$, and $g_s\to A_*$.
Substituting these limits into Eq.~\eqref{eq:satt_closed_balanced}, every numerator and denominator term carries a common factor $2s^2$, which cancels.
The resulting limit is
\begin{equation}
\nu_\Delta(0^+)
=
\frac{A_*^2}
{\frac{c^2}{m^2}\left(\frac{1}{L-1}+\frac{1}{d_W}\right)
-\frac{2c}{m d_W}A_*
+\frac{A_*^2}{d_W}},
\qquad c=\frac{L-1}{L}.
\label{eq:satt_boundary_limit}
\end{equation}
For the common intercept-plus-slope design,
\[
A_* = \frac{1}{L}\left\{\frac{1}{m}+\frac{(t_* - \bar t)^2}{\sum_{j=1}^m(t_j-\bar t)^2}\right\}.
\]
Thus $A_*$ is minimized for prediction points near the time-design centroid.

The denominator in Eq.~\eqref{eq:satt_boundary_limit} has a design-only term, a term linear in $A_*$, and a term quadratic in $A_*$. To display these pieces, write it as
\[
K_0-K_1A_*+\frac{A_*^2}{d_W},
\qquad
K_0=\frac{c^2}{m^2}
\left(\frac{1}{L-1}+\frac{1}{d_W}\right),
\qquad
K_1=\frac{2c}{m d_W}.
\]
The constant $K_0$ depends only on the design $(L,m,d_W)$, while the signed cross term $-K_1A_*$ and the quadratic term $A_*^2/d_W$ vary with the fixed-effect leverage $A_*$. This decomposition gives two useful limiting cases. If $A_*$ is large enough that the quadratic term $A_*^2/d_W$ dominates both $K_0$ and the magnitude of the linear term $K_1A_*$, then $\nu_\Delta(0^+)\approx d_W$, the within-lot residual degrees of freedom. If the prediction point is near the time-design centroid, then $A_*$ is near its minimum and $K_0$ dominates the denominator. In that case, $\nu_\Delta(0^+)\approx A_*^2/K_0$, which can be very small.

The worked example is in the small-$A_*$ case: $t_*=24$ is close to $\bar t=22$, so $A_*=0.008014$ and Eq.~\eqref{eq:satt_boundary_limit} gives $\nu_\Delta(0^+)\approx0.071$. For $L=14$, $m=9$, and $d_W=111$, the three denominator pieces $\{K_0,\,-K_1A_*,\,A_*^2/d_W\}$ evaluate to $9.15\times10^{-4}$, $-1.49\times10^{-5}$, and $5.79\times10^{-7}$, respectively, confirming that $K_0$ dominates near the time-design centroid. Thus predictions near the time-design centroid can drive the right-hand delta-method DDF toward zero when the random-intercept structure is retained with a small positive variance estimate. Consequently, apparent middle-of-design widening (Figure~\ref{fig:sas-overlay}) is not caused by a large prediction standard error. It is caused by the Satterthwaite DDF collapsing where the fixed-effect leverage term $A_*$ is smallest, so the inflated $t$ multiplier can dominate the smaller standard error.

\newpage
\section{Additional sensitivity analyses}
\label{app:additional_sensitivity}

This appendix reports additional analyses that support the main results: Kenward--Roger mechanism plots, 5/10/20\% threshold sensitivity for SAT\_reduced, the SAT\_covtest25 p-value reduction comparator, and matched Model~(1) simulations with nonzero random-slope variance.

\subsection{Kenward--Roger mechanism plots}
\label{app:kr_sensitivity}

\begin{figure}[htbp]
\centering
\includegraphics[width=0.82\linewidth]{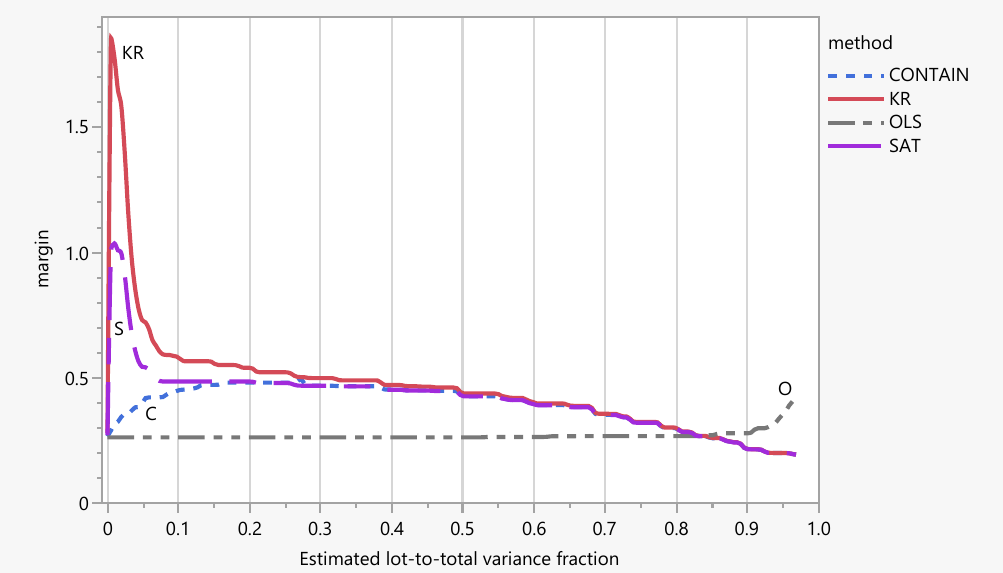}
\caption{Additional Kenward--Roger mechanism analysis: mean lower-limit margin by fitted lot variance fraction.}
\label{fig:app_kr_margin_bin}
\end{figure}

\begin{figure}[htbp]
\centering
\includegraphics[width=0.82\linewidth]{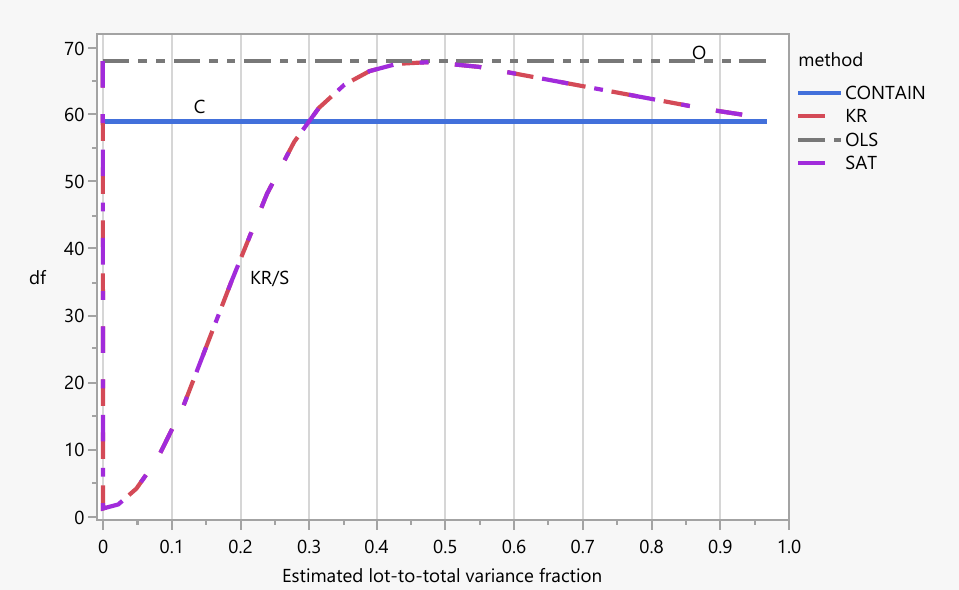}
\caption{Additional Kenward--Roger mechanism analysis: mean denominator degrees of freedom by fitted lot variance fraction.}
\label{fig:app_kr_df_bin}
\end{figure}

Figures~\ref{fig:app_kr_margin_bin} and~\ref{fig:app_kr_df_bin} repeat the mechanism summaries with KR added. KR follows the same near-boundary DDF-collapse pattern as SAT. At small positive fitted variance-fraction values, SAT and KR have similarly low DDF, and the mean lower-limit margin is larger for KR than for SAT because of the KR standard-error adjustment; containment remains stable.

\subsection{Threshold sensitivity for SAT\_reduced}
\label{app:threshold_sensitivity}

\begin{figure}[htbp]
\centering
\includegraphics[width=0.82\linewidth]{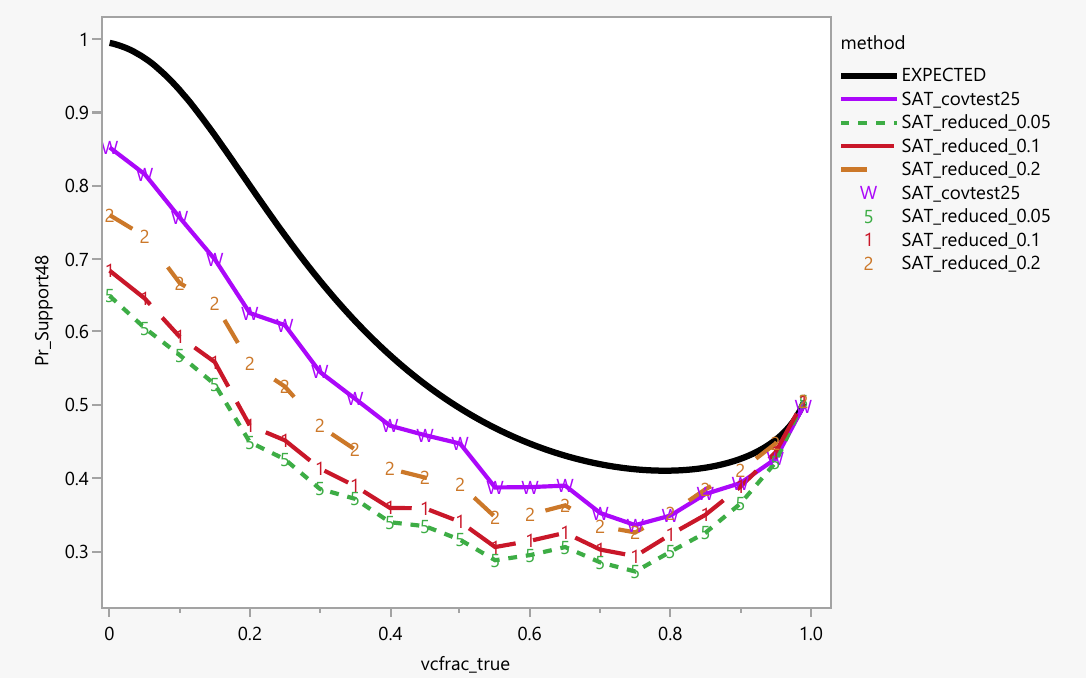}
\caption{SAT\_reduced threshold sensitivity for $\Pr(\mathrm{Support48})$ under the baseline 57-month calibration, with SAT\_covtest25 included as the p-value reduction comparator.}
\label{fig:app_threshold_support}
\end{figure}

\begin{figure}[htbp]
\centering
\includegraphics[width=0.82\linewidth]{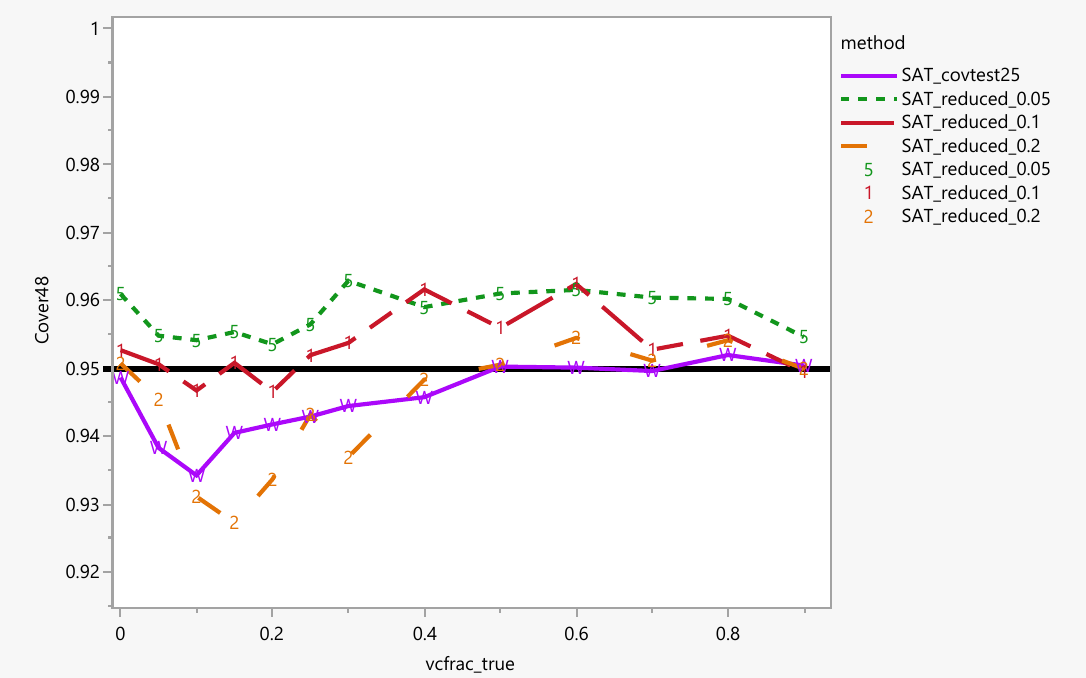}
\caption{SAT\_reduced threshold sensitivity for pointwise coverage at $t_*=48$ months, with SAT\_covtest25 included as the p-value reduction comparator.}
\label{fig:app_threshold_coverage}
\end{figure}

Figures~\ref{fig:app_threshold_support} and~\ref{fig:app_threshold_coverage} compare 5\%, 10\%, and 20\% variance-contribution thresholds. Support48 increases as the threshold becomes more aggressive: the maximum Support48 is 0.649 for 5\%, 0.684 for 10\%, and 0.760 for 20\%. Coverage remains conservative or close to nominal for 5\% and 10\%, whereas the 20\% threshold has lower coverage in the small-to-moderate variance region, with a minimum of 0.927. These results support keeping 10\% as the primary operational rule rather than adopting the more aggressive 20\% threshold.

\subsection{SAT\_covtest25 p-value reduction comparator}
\label{app:covtest_sensitivity}

\begin{figure}[htbp]
\centering
\includegraphics[width=0.86\linewidth]{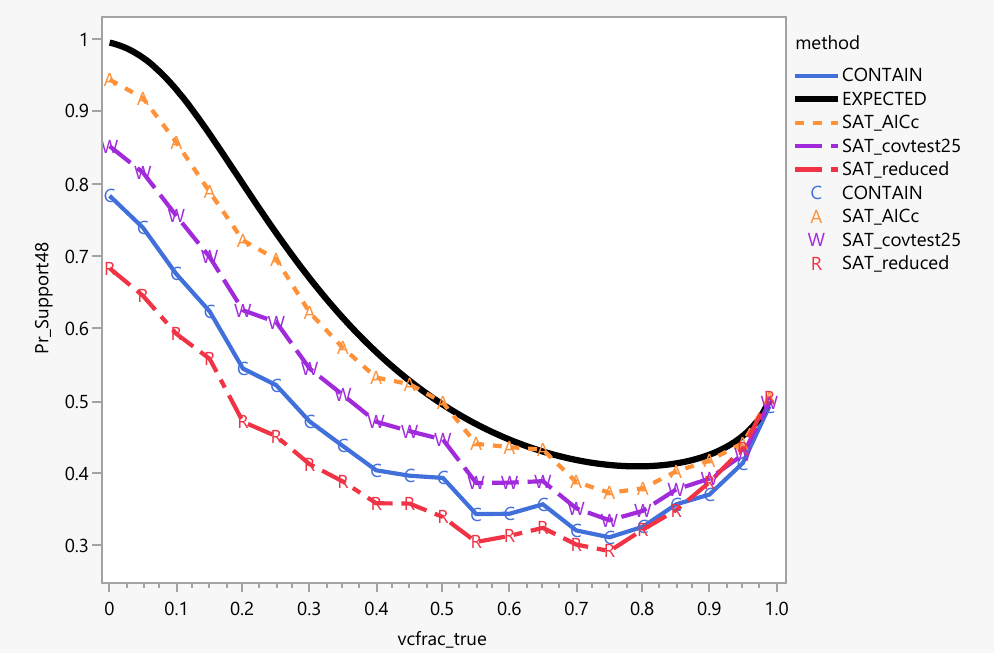}
\caption{SAT\_covtest25 Support48 sensitivity under the baseline 57-month calibration, with SAT\_reduced, CONTAIN, SAT\_AICc, and the known-variance reference curve shown for reference.}
\label{fig:app_covtest_support}
\end{figure}

\begin{figure}[htbp]
\centering
\includegraphics[width=0.86\linewidth]{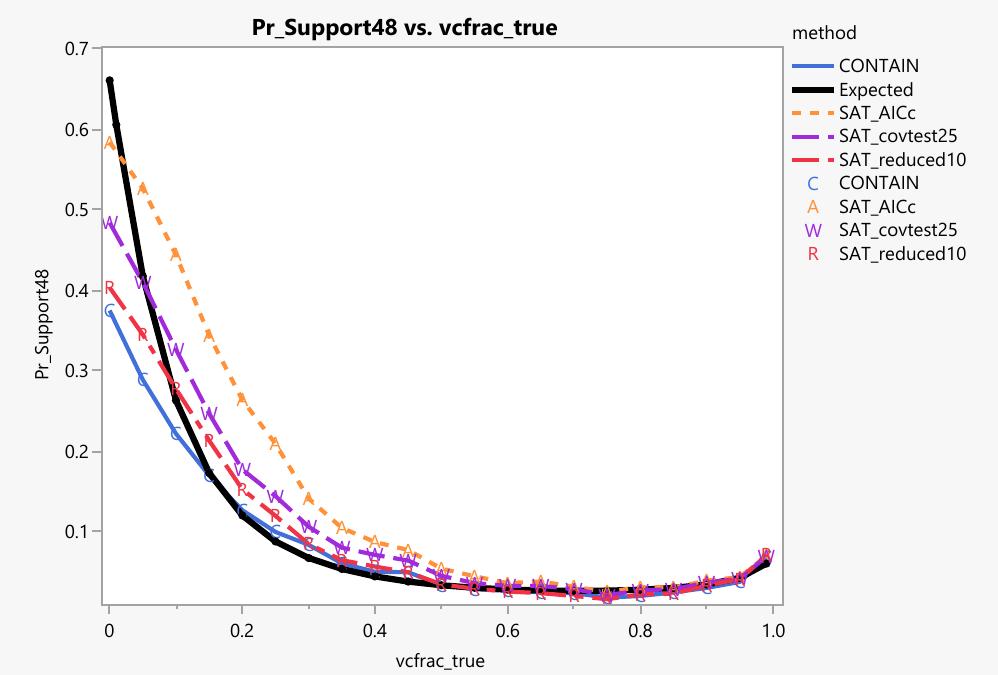}
\caption{SAT\_covtest25 Support48 sensitivity under the tighter 52-month calibration, with SAT\_reduced, CONTAIN, SAT\_AICc, and the known-variance reference curve shown for reference.}
\label{fig:app_covtest_support_52mo}
\end{figure}

SAT\_covtest25 uses raw \texttt{PROC MIXED} \texttt{COVTEST} p-values in the same conditional Model~(1) $\rightarrow$ Model~(2) $\rightarrow$ Model~(3) step-down sequence described above, removing random-effect terms when the variance-component p-value exceeds 0.25. This is a natural Q1E-style comparator, but it is not a primary recommendation because testing variance components at zero is non-regular. Figures~\ref{fig:app_covtest_support} and~\ref{fig:app_covtest_support_52mo} summarize its Support48 behavior relative to SAT\_reduced, CONTAIN, SAT\_AICc, and the known-variance reference curve; Figure~\ref{fig:app_threshold_coverage} includes the corresponding baseline pointwise-coverage comparison. Under the baseline 57-month calibration, SAT\_covtest25 increases Support48 relative to SAT\_reduced in low-variance settings but remains below the known-variance reference curve. Under the tighter 52-month calibration, SAT\_covtest25 is less aggressive than SAT\_AICc but can exceed the known-variance reference curve in the small-to-moderate variance range. For example, at $\mathrm{vcfrac}_{\mathrm{true}}=0.10$, Support48 is 0.326 for SAT\_covtest25, 0.277 for SAT\_reduced, 0.443 for SAT\_AICc, and 0.263 for the reference curve. This added support coincides with lower pointwise coverage in the baseline coverage diagnostic: SAT\_covtest25 has minimum coverage 0.934 versus 0.947 for SAT\_reduced.

\subsection{Matched Model~(1) data-generating simulations}
\label{app:m1_sensitivity}

\begin{figure}[htbp]
\centering
\includegraphics[width=0.90\linewidth]{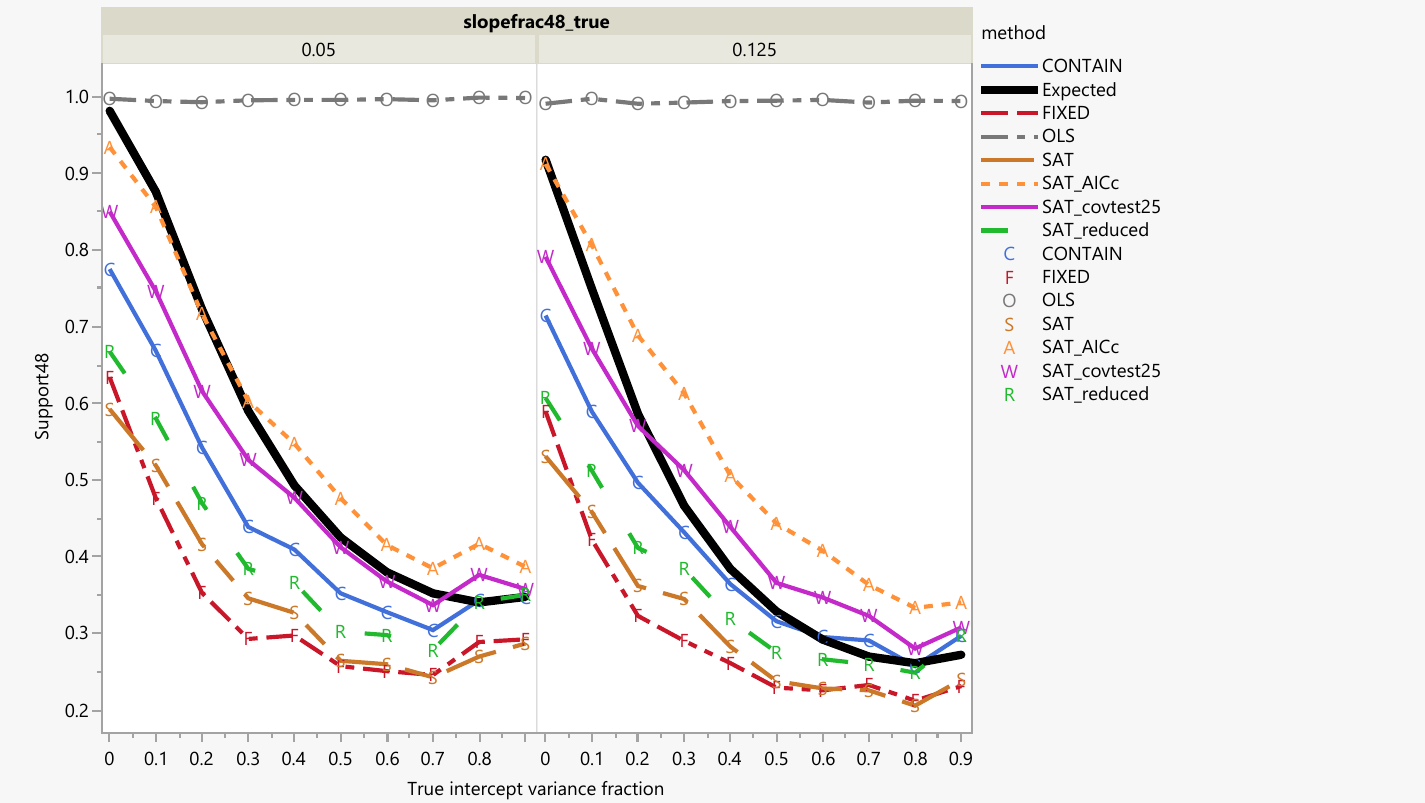}
\caption{Matched Model~(1) sensitivity: $\Pr(\mathrm{Support48})$ by random-intercept variance fraction, with panels for random-slope contribution at $t_*=48$ months. The Expected curve is the known-variance reference computed under the corresponding Model~(1) data-generating parameters.}
\label{fig:app_m1_support}
\end{figure}

\begin{figure}[htbp]
\centering
\includegraphics[width=0.90\linewidth]{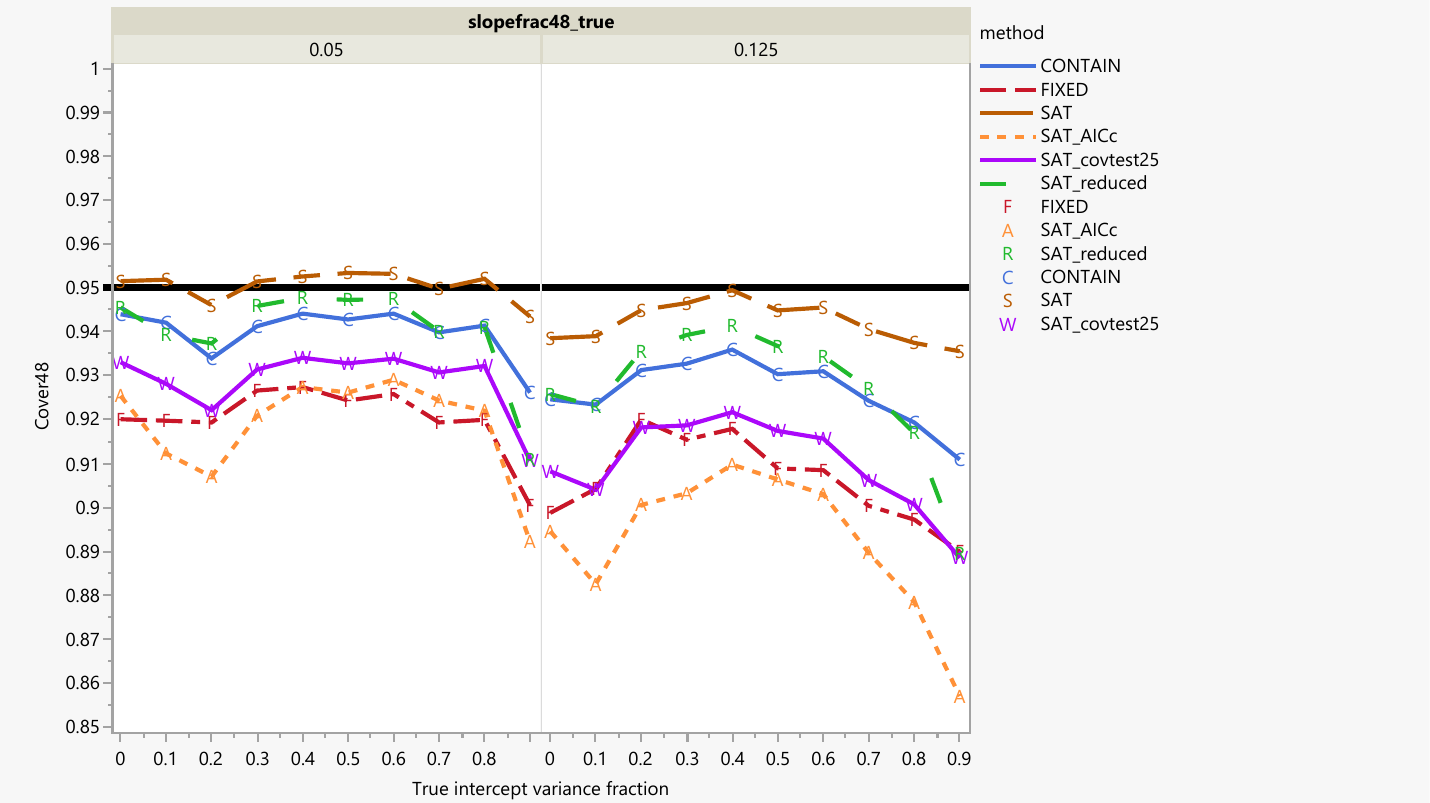}
\caption{Matched Model~(1) sensitivity: pointwise coverage at $t_*=48$ months by random-intercept variance fraction, with panels for random-slope contribution at $t_*=48$ months.}
\label{fig:app_m1_coverage}
\end{figure}

Figures~\ref{fig:app_m1_support} and~\ref{fig:app_m1_coverage} show the matched Model~(1) sensitivity results. In these simulations the intercept/residual base scale is the same as in the main simulation grid: $\sigma_{b0}^2=\mathrm{vcfrac}_{\mathrm{true}}$ and $\sigma_e^2=1-\mathrm{vcfrac}_{\mathrm{true}}$, so $\sigma_{b0}^2+\sigma_e^2=1$. The random-slope variance is then added separately and chosen to give a specified slope contribution to the marginal variance at $t_*=48$ months:
\[
p_{b1,\mathrm{true}}(48)
=\frac{48^2\sigma_{b1}^2}{\sigma_{b0}^2+48^2\sigma_{b1}^2+\sigma_e^2}
=\frac{48^2\sigma_{b1}^2}{1+48^2\sigma_{b1}^2}.
\]
Solving gives $\sigma_{b1}^2=\{p_{b1,\mathrm{true}}(48)/(1-p_{b1,\mathrm{true}}(48))\}/48^2$. The two panels correspond to $p_{b1,\mathrm{true}}(48)=0.05$ and 0.125, giving slope variances $2.28\times10^{-5}$ and $6.20\times10^{-5}$.
The known-variance reference curve extends the Appendix~\ref{app:baselinepass_lsl} calculation to the diagonal random-intercept/random-slope covariance structure used in these simulations.

The main qualitative conclusion is a trade-off rather than dominance by a single method. SAT has the most stable pointwise coverage (0.936--0.953) and is generally the most conservative mixed-model procedure for Support48. CONTAIN increases Support48 relative to SAT without data-dependent model switching, but its pointwise coverage is modestly below nominal in the moderate-slope/high-intercept-variance corner (minimum 0.911). Even in this under-nominal corner, the pointwise coverage for CONTAIN and SAT\_reduced remains at or above that of the Q1E-style FIXED comparator over nearly all matched Model~(1) settings, so the modest undercoverage should be interpreted relative to an accepted fixed-lot benchmark as well as relative to the nominal 0.95 line. AICc and OLS increase Support48 in some regions but can fall below nominal coverage when random slopes are truly present; SAT\_AICc has minimum coverage 0.857. SAT\_covtest25 generally sits between SAT\_reduced and SAT\_AICc in this trade-off: it raises Support48 above the 10\% variance-contribution rule in low-variance settings but remains less aggressive than AICc. Its pointwise coverage drops to 0.889 in the moderate-slope/high-intercept-variance corner (Figure~\ref{fig:app_m1_coverage}). 

\clearpage

\bibliographystyle{IEEEtran}

\bibliography{references_revised_23MAY}

@article{BurdickErmer2019,
  author  = {Burdick, Richard K. and Ermer, Joachim},
  title   = {Precision of the reportable value---Statistical optimization of the number of replicates},
  journal = {Journal of Pharmaceutical and Biomedical Analysis},
  year    = {2019},
  volume  = {162},
  pages   = {149--157},
  doi     = {10.1016/j.jpba.2018.08.062},
  url     = {https://doi.org/10.1016/j.jpba.2018.08.062}
}

@article{ChatterjeeLahiriLi2008,
  author  = {Chatterjee, Snigdhansu and Lahiri, Partha and Li, Huilin},
  title   = {Parametric bootstrap approximation to the distribution of {EBLUP} and related prediction intervals in linear mixed models},
  journal = {The Annals of Statistics},
  year    = {2008},
  volume  = {36},
  number  = {3},
  pages   = {1221--1245},
  doi     = {10.1214/07-AOS512},
  url     = {https://doi.org/10.1214/07-AOS512}
}

@article{selfliang,
  author  = {Self, Steven G. and Liang, Kung-Yee},
  title   = {Asymptotic Properties of Maximum Likelihood Estimators and Likelihood Ratio Tests Under Nonstandard Conditions},
  journal = {Journal of the American Statistical Association},
  year    = {1987},
  volume  = {82},
  number  = {398},
  pages   = {605--610},
  doi     = {10.2307/2289471},
  url     = {https://doi.org/10.2307/2289471}
}

@article{Sugiura1978,
  author  = {Sugiura, Nariaki},
  title   = {Further analysis of the data by {Akaike}'s information criterion and the finite corrections},
  journal = {Communications in Statistics - Theory and Methods},
  year    = {1978},
  volume  = {7},
  number  = {1},
  pages   = {13--26},
  doi     = {10.1080/03610927808827599},
  url     = {https://doi.org/10.1080/03610927808827599}
}

@article{HurvichTsai1989,
  author  = {Hurvich, Clifford M. and Tsai, Chih-Ling},
  title   = {Regression and time series model selection in small samples},
  journal = {Biometrika},
  year    = {1989},
  volume  = {76},
  number  = {2},
  pages   = {297--307},
  doi     = {10.1093/biomet/76.2.297},
  url     = {https://doi.org/10.1093/biomet/76.2.297}
}

@techreport{ICHQ1Draft2025,
  author      = {{International Council for Harmonisation of Technical Requirements for Pharmaceuticals for Human Use}},
  title       = {{ICH} Harmonised Guideline: Stability Testing of Drug Substances and Drug Products, {Q1}},
  institution = {International Council for Harmonisation of Technical Requirements for Pharmaceuticals for Human Use},
  year        = {2025},
  type        = {Draft guideline},
  number      = {Q1},
  note        = {Step 2b draft version; endorsed 11 April 2025; released for public consultation in 2025},
  url         = {https://database.ich.org/sites/default/files/ICH_Q1EWG_Step2_Draft_Guideline_2025_0411.pdf},
  urldate     = {2026-05-23}
}

@techreport{ICHQ1E2003,
  author      = {{International Council for Harmonisation of Technical Requirements for Pharmaceuticals for Human Use}},
  title       = {{ICH} Harmonised Tripartite Guideline: Evaluation for Stability Data, {Q1E}},
  institution = {International Council for Harmonisation of Technical Requirements for Pharmaceuticals for Human Use},
  year        = {2003},
  type        = {Guideline},
  number      = {Q1E},
  note        = {Current Step 4 version; dated 6 February 2003},
  url         = {https://database.ich.org/sites/default/files/Q1E_Guideline.pdf},
  urldate     = {2026-05-23}
}

@manual{JMPMixedModelOptions19,
  author       = {{JMP Statistical Discovery LLC}},
  title        = {{JMP Pro 19.1}: Mixed Model Report and Options},
  organization = {JMP Statistical Discovery LLC},
  year         = {2025},
  note         = {JMP Help, Version 19.1},
  url          = {https://www.jmp.com/support/help/en/19.1/#page/jmp/mixed-model-report-and-options.shtml},
  urldate      = {2026-05-23}
}

@article{henderson1959,
  author  = {Henderson, C. R. and Kempthorne, Oscar and Searle, S. R. and von Krosigk, C. M.},
  title   = {The Estimation of Environmental and Genetic Trends from Records Subject to Culling},
  journal = {Biometrics},
  year    = {1959},
  volume  = {15},
  number  = {2},
  pages   = {192--218},
  doi     = {10.2307/2527669},
  url     = {https://doi.org/10.2307/2527669}
}

@misc{KarlMendeleyData2025,
  author    = {Karl, Andrew},
  title     = {Simulation Code for Conditional-Mean Inference in Random-Lot Stability Models},
  year      = {2026},
  publisher = {Mendeley Data},
  version   = {V4},
  doi       = {10.17632/4hcmr7gfcn.4},
  url       = {https://doi.org/10.17632/4hcmr7gfcn.4},
  urldate   = {2026-05-31}
}

@article{KenwardRoger1997,
  author  = {Kenward, Michael G. and Roger, James H.},
  title   = {Small sample inference for fixed effects from restricted maximum likelihood},
  journal = {Biometrics},
  year    = {1997},
  volume  = {53},
  number  = {3},
  pages   = {983--997},
  doi     = {10.2307/2533558},
  url     = {https://doi.org/10.2307/2533558}
}

@article{KenwardRoger2009,
  author  = {Kenward, Michael G. and Roger, James H.},
  title   = {An improved approximation to the precision of fixed effects from restricted maximum likelihood},
  journal = {Computational Statistics \& Data Analysis},
  year    = {2009},
  volume  = {53},
  number  = {7},
  pages   = {2583--2595},
  doi     = {10.1016/j.csda.2008.12.013},
  url     = {https://doi.org/10.1016/j.csda.2008.12.013}
}

@manual{SASMixedDDFM152,
  author       = {{SAS Institute Inc.}},
  title        = {{SAS/STAT} User's Guide: The {MIXED} Procedure, {MODEL} Statement},
  organization = {SAS Institute Inc.},
  year         = {2026},
  note         = {SAS Viya Platform Programming Documentation, 2026.05; includes the {DDFM=} denominator degrees-of-freedom option},
  url          = {https://documentation.sas.com/doc/en/pgmsascdc/v_075/statug/statug_mixed_syntax10.htm},
  urldate      = {2026-05-23}
}

@article{Satterthwaite1946,
  author  = {Satterthwaite, Franklin E.},
  title   = {An approximate distribution of estimates of variance components},
  journal = {Biometrics Bulletin},
  year    = {1946},
  volume  = {2},
  number  = {6},
  pages   = {110--114},
  doi     = {10.2307/3002019},
  url     = {https://doi.org/10.2307/3002019}
}

@article{Scheipl2008,
  author  = {Scheipl, Fabian and Greven, Sonja and K{\"u}chenhoff, Helmut},
  title   = {Size and power of tests for a zero random effect variance or polynomial regression in additive and linear mixed models},
  journal = {Computational Statistics \& Data Analysis},
  year    = {2008},
  volume  = {52},
  number  = {7},
  pages   = {3283--3299},
  doi     = {10.1016/j.csda.2007.10.022},
  url     = {https://doi.org/10.1016/j.csda.2007.10.022}
}

@book{mvtnormManual,
  author    = {Genz, Alan and Bretz, Frank},
  title     = {Computation of Multivariate Normal and t Probabilities},
  series    = {Lecture Notes in Statistics},
  volume    = {195},
  publisher = {Springer-Verlag},
  address   = {Heidelberg},
  year      = {2009},
  isbn      = {978-3-642-01688-2},
  doi       = {10.1007/978-3-642-01689-9},
  url       = {https://doi.org/10.1007/978-3-642-01689-9}
}

@incollection{Pack2018Stability,
  author    = {Pack, Laura D.},
  title     = {Statistical Analysis of Stability Studies},
  booktitle = {Statistics for Biotechnology Process Development},
  editor    = {Coffey, Todd and Yang, Harry},
  edition   = {1},
  publisher = {Chapman and Hall/CRC},
  address   = {Boca Raton, FL},
  year      = {2018},
  chapter   = {7},
  pages     = {169--252},
  isbn      = {9781315120034},
  doi       = {10.1201/9781315120034-7},
  url       = {https://doi.org/10.1201/9781315120034-7}
}

@manual{JMPKH,
  author       = {{JMP Statistical Discovery LLC}},
  title        = {{JMP Pro 19.1}: Statistical Details for the Kackar--Harville Correction},
  organization = {JMP Statistical Discovery LLC},
  year         = {2025},
  note         = {JMP Help, Version 19.1},
  url          = {https://www.jmp.com/support/help/en/19.1/#page/jmp/statistical-details-for-the-kackarharville-correction-2.shtml},
  urldate      = {2026-05-23}
}

@article{GiesbrechtBurns1985,
  author  = {Giesbrecht, F. G. and Burns, J. C.},
  title   = {Two-stage analysis based on a mixed model: large-sample asymptotic theory and small-sample simulation results},
  journal = {Biometrics},
  year    = {1985},
  volume  = {41},
  number  = {2},
  pages   = {477--486},
  doi     = {10.2307/2530872},
  url     = {https://doi.org/10.2307/2530872}
}

@article{robinson1991blup,
  author  = {Robinson, George K.},
  title   = {That {BLUP} is a Good Thing: The Estimation of Random Effects},
  journal = {Statistical Science},
  year    = {1991},
  volume  = {6},
  number  = {1},
  pages   = {15--32},
  doi     = {10.1214/ss/1177011926},
  url     = {https://doi.org/10.1214/ss/1177011926}
}

@book{SearleCasellaMcCulloch1992,
  author    = {Searle, Shayle R. and Casella, George and McCulloch, Charles E.},
  title     = {Variance Components},
  publisher = {John Wiley \& Sons},
  address   = {New York, NY},
  year      = {1992}
}

\end{document}